\documentclass[preprint]{aastex}
\usepackage{emulateapj5,apjfonts,psfig}

\begin{document}
\title{Dust and the Infrared Kinematic\\
Properties of Early-Type Galaxies}
\author{Julia D. Silge and Karl Gebhardt}
\affil{Astronomy Department, University of Texas at Austin,
    1 University Station C1400, Austin, TX 78712}
\email{dorothea@astro.as.utexas.edu, gebhardt@astro.as.utexas.edu}

\begin{abstract}

We have obtained spectra and measured the stellar kinematics in a
sample of 25 nearby early-type galaxies (with velocity dispersions
from less than 100 km s$^{-1}$ to over 300  km s$^{-1}$) using the 
near-infrared CO absorption bandhead at 2.29 $\mu$m. Our median
uncertainty for the dispersions is $\sim$10\%.
We examine the effects of dust on existing optical kinematic 
measurements. We find that the near-infrared velocity dispersions 
are in general smaller than optical velocity dispersions, with 
differences as large as 30\%.  The median difference is 11\% 
smaller, and the effect is of greater magnitude for higher 
dispersion galaxies.  The lenticular galaxies 
(18 out of 25) appear to be
causing the shift to lower dispersions while the classical
ellipticals (7 out of 25) 
are consistent between the two wavelength regimes.
If uniformly distributed dust causes these differences, 
we would expect to find a correlation between the relative
amount of dust in a galaxy and the fractional change in
dispersion, but we do not find such a correlation.  We do see
correlations both between velocity dispersion and CO
bandhead equivalent width, and velocity dispersion and Mg$_{2}$
index. The differences in dispersion are not well explained by 
current models of dust absorption.  The lack of correlation
between the relative amount of dust and shift in dispersion
possibly suggets that dust does not have a similar distribution
from galaxy to galaxy. The CO equivalent widths of 
these galaxies are quite high ($\ga$ 10 \AA~for almost all), 
requiring the light at these wavelengths to be dominated by
very cool stars.

\end{abstract}

\section{Introduction}

Before infrared detectors became available,
the amount of dust in a galaxy was estimated from patchy,
optically visible obscuration.  However, since dust emits in the far
infrared, observing in that part of the spectrum gives a
better measure of the total amount of dust. 
Recent far-infrared observations have shown 
that galaxies, even ellipticals and bulges of galaxies, contain 
an unexpectedly large amount of dust \citep{gou95}.  This dust may bias the
optical region by skewing photometric and/or kinematic data through 
absorption and scattering.  Observing in
the near infrared allows us to minimize these problems.
Many groups have studied the photometric bias introduced by
dust, and here we study its kinematic effect.
Previous attempts at studying kinematic dust effects were
limited to those galaxies with obvious concerns.
There has been no systematic investigation based on a general
galaxy population.  Our goal here is to quantify these effects
by studying a subset of galaxies that are thought to have
minimal dust problems: early-type galaxies.
Also, observing at these longer wavelengths traces
the older, redder stellar population and minimizes effects due to
recent star formation.  Thus, kinematics in this spectral regime
should produce the best measure of the underlying stellar
potential of the galaxy.  As infrared instrumentation and telescopes
become more efficient, this region is quickly becoming very
important.  As a first step, we must characterize
the IR features for kinematic analysis.  Furthermore,
analysis of these possible biases and problems is important
to accurately interpret the data we will soon have from very large
optical samples.

Observations using 
the Infrared Astronomical Satellite (IRAS) have shown that 
galaxies, even those previously thought to be practically
dust-free such as
ellipticals and bulges of galaxies, can contain a
large amount of dust.  Dust masses determined from
far infrared IRAS flux densities are about a factor of 10
higher than those estimated from optically visible patches and lanes
\citep{gou95}. More sophisticated estimates of the dust mass
using submillimeter data give values even higher \citep{kwa92,wik95}.
There is much more dust out there than
is apparent when observing in the visible.  \citet{gou95}
explain this by suggesting that most of this dust exists as a diffusely
distributed component, which would be undetectable in the optical regime.

How does dust affect our measurements of galaxies?
Models of elliptical
galaxies considering only dust absorption show effects on both
the photometric and kinematic data.  The photometric effects include
a global attenuation,
strong extinction in the central regions, and an increasing apparent
core radius \citep{bae00a}. The core radius is one of the
primary parameters used to study galaxy correlation functions
and according to these models appears larger than its true value 
because the luminosity profile is flattened towards the center by
the dust. The core radius can increase by over 25\% for an
optical depth $\tau \approx  2$ (a moderate amount of dust), 
an important effect
considering the small amount of scatter in the
galaxy correlation functions. 
The effects of absorption on kinematic properties \citep{bae00b}, 
calculated by building semi-analytic
dusty galaxy models and then modeling the
synthetic data assuming no dust,
depend on the shape of the velocity dispersion tensor of the input model.  
For radial and isotropic
orbital structures, the inferred dynamical mass is significantly underestimated
while the inferred orbital structure is mostly unaffected.
For an optical depth $\tau \approx 2$, the dynamical mass appears 
20\% smaller.
For galaxies with tangential orbital
structures, the dynamical mass is not affected much 
but the inferred orbital structure
appears more radial, even for small amounts of dust. 
Both these effects are due to dust
preferentially obscuring light from high-velocity regions of the galaxies.
\citet{bae00b} find that dust absorption does not significantly affect
the velocity dispersion but that the dynamical structure is not correctly
recovered.
\citet{bae01, bae02} construct models with both dust absorption and dust
scattering (using Monte Carlo methods),
and find different effects on the observed
kinematics which still depend on the orbital structure of the
input galaxy.
At small radii, dust causes the central
dispersion to appear smaller for radial and isotropic galaxies and to 
appear larger for tangential galaxies.  This effect is small; an optical
depth $\tau \approx 2$ causes a change of a few percent.  There are
dramatic changes at large radii, however. The attenuation by dust, mostly
the scattering, results in high-velocity wings in the line-of-sight 
velocity distribution (LOSVD) in the 
outer parts of the galaxy. For $\tau \approx 1$, the projected dispersion
at large radii can increase by over 40\%.
Both these effects (at small and large radii) are caused by the scattering
of light from high-velocity stars into lines of sight at which such
stars do not exist.

Galaxies which are visibly dusty pose a less subtle problem. Although they,
too, may have a diffusely distributed gas component, the high level of
obscuration of visible light from patches and lanes hinders optical
spectroscopy of the galaxies.  There are a significant number of galaxies
out there for which the existing kinematic data are suspect and untrustworthy
because of visible dust.  Dust lanes are already seen in about half
of all ellipticals \citep{kor89,lau95,van95,res01,tra01}, 
with higher detection rates as
search techniques (especially spatial resolution) improve. 
These galaxies are often not included in
kinematic studies because data for them are unrealiable, introducing
potential bias into current kinematic samples. It
is important to observationally constrain how observed galaxy kinematics
are being affected or biased by interstellar dust.

Galaxy kinematics are measured in the optical using several
different spectral features.  The most common include 
the H and K lines (Ca II) near 4000 \AA, the Mg $b$ lines 
near 5175 \AA, and the Ca II triplet near 8500 \AA.
Studies comparing kinematic measurements made using different
features, such as \citet{nel95}, do not find significant
differences between them, although \citet{bar02} find the 
Mg $b$ lines are sensitive to template mismatch and the details
of the fitting procedure while the Ca II triplet is much more
robust.  All these spectral features may be 
affected by dust to some extent.
Dust is very opaque in the $B$ band where the H and K lines
are located and absorbs less efficiently at longer
wavelengths.  At 5000 \AA, dust absorption is 75\% of that in $B$ and
decreases to 40\% in $I$ where the Ca II triplet is located.  Moving
to a feature like the Ca II triplet certainly reduces problems
associated with extinction, but effects from dust can be further
minimized by moving to longer wavelengths. The extinction in the 
$K$ band is only 7\% of that at $B$ \citep{gaf95,bae02}.

In this project, we measure the stellar kinematics in a sample of 
25 nearby early-type galaxies using the $K$-band CO absorption bandhead.
Using wavelengths this long will allow us to avoid dust absorption
and gain a more accurate understanding of the kinematics of these
galaxies.
This sample will provide a check of possible
kinematic biases using optical light.
Calibrating the CO bandhead for 
galaxy kinematic measurements is an important goal of this study.
Working in the near infrared holds great
promise for both scientific motivations and future instrumentation.
Specifically, adaptive optics, in which the optical system is
continually adjusted to compensate for the effects of seeing, work
best in the near infrared.  The longer wavelengths in this
part of the spectrum allow less stringent requirements on optics
adjustment.  Using the CO bandhead will optimize kinematic analysis
for adaptive optics.

This technique can also be extended to other scientific questions.
Our current understanding of the kinematics of galaxies has led to the
study of relations between different characteristics of galaxies in
$\kappa$-space (i.e. Fundamental Plane, Faber-Jackson relation,
Tully-Fisher relation; e.g.  Burstein et al.\ 1997).  
These relations have been
derived using data from optical stellar emission lines, so 
dusty, complicated, messy galaxies are not included in the analyses.
The galaxy sample itself may be biased because of the exclusion of these
galaxies.  Also,
if bulges and elliptical galaxies have a diffusely distributed dust
component, these
analyses may suffer from important problems due to internal absorption
and extinction. 
Observing in the $K$ band would avoid these problems.

This paper is organized as follows: Section 2 presents the data,
Section 3 discusses our analysis techniques, Section 4 presents
our results for the infrared kinematic measurements and how these
results compare to other data for these galaxies, and Section 5
summarizes.

\section{Data}

\subsection{The Sample}

Table \ref{tbl-1} shows the 25 galaxies in this project.
The sample contains ellipticals
and lenticulars, and has redshifts less than 5000 km s$^{-1}$.
Our sample is made largely of galaxies from \citet{ton01}.  
\citet{ton01} present distance moduli for 300 nearby
galaxies from surface brightness fluctuations.  This presents an ideal
sample to draw from for this study; these galaxies are
systems with bulges (mostly E and S0, with some S), are essentially 
complete to a redshift of 1000 km s$^{-1}$, and have well-determined
distances (which are necessary to calculate the dust mass and other 
physical properties).
The galaxy type and heliocentric velocity (from NED), distance modulus
(from Tonry et al.\ 2001), and the calculated distance are listed in 
Table \ref{tbl-1}.
The sample contains about twice as many S0 galaxies as classical
ellipticals, which is reflective of the galaxies in the SBF sample at
these redshifts.
For the few galaxies in this sample that are not in \citet{ton01},
the distance is calculated from the recessional velocity
using $H_0=70$ km s$^{-1}$ Mpc$^{-1}$.
This is relatively accurate
because the few galaxies without distance moduli
have redshifts $>$ 2000 km s$^{-1}$.
These galaxies have a range of IRAS dust characteristics, from
no detectable dust to several million solar masses of dust 
(see Section 3.3).
All galaxies have associated IRAS fluxes from \citet{kna89}
and the 1994 correction to those data (available at 
\url{http://cdsweb.u-strasbg.fr/CDS.html}).

\subsection{The CO Bandhead}

The 2.29 $\mu$m (2-0) $^{12}$CO absorption bandhead 
from evolved red stars is the strongest absorption feature
in galactic spectra in the 1--3 $\mu$m range.
This is the optimal 
range for studying stellar kinematics  because it is long enough 
to  minimize 
extinction from dust but short enough to avoid dilution of
the stellar continuum by hot dust \citep{gaf95}. The feature is in a dark
part of the infrared sky spectrum and is intrinsically sharp and deep,
making it very sensitive to stellar motions \citep{les94}.

The CO bandhead is present
in late-type stars, increasing in strength with decreasing
effective temperature or increasing radius.
The CO bandhead has been used to measure the stellar kinematics
of galaxies in recent years, but only in galaxies such as starbursts
and mergers where optical kinematic measurements are seriously hindered 
\citep{tam91,gaf93,doy94,les94,shi94,shi96,pux97}.
There are no velocity dispersion measurements from the CO bandhead
for galaxies which are well-studied in the optical, and thus there
is no information on possible differences between optical and infrared
kinematics.

\subsection{Observations and Data Reduction}

Observations were taken during 27 nights in six observing runs 
between
December 2000 and January 2002. We use CoolSpec \citep{les00}, 
a near-infrared
grating spectrometer, on the 2.7-m telescope at McDonald
Observatory to measure the stellar kinematics in our sample.
CoolSpec has a 256 $\times$ 256 HgCdTe NICMOS III detector array.
Using a 240 l/mm grating and 1.8$^{\prime\prime}$ $\times$ 
90$^{\prime\prime}$ slit, our 
spectral resolution is 2300, measured from calibration lamp lines.
This gives a full width half maximum
(FWHM) resolution of approximately
130 km s$^{-1}$, which allows us to study galaxies with velocity 
dispersions down to approximately 50 km s$^{-1}$.  Resolving the 
dispersions of early-type galaxies is easily within reach of this
observational set-up. 
We have obtained spectra for about 40 galaxies, 27 of which are early-type.
We have successfully extracted the line-of-sight velocity distribution
for 25 of these galaxies, which make up the sample for this project.
The other two galaxies were too large on the chip to allow for
good sky subtraction.

The spatial and spectral scale are 0$^{\prime\prime}$.35
pixel$^{-1}$ and 24.6 km s$^{-1}$
pixel$^{-1}$, respectively. The latter gives a spectral range of just
under 0.05 $\mu$m, which is large enough to provide good coverage of the CO
bandhead and continuum on both sides.  Technically, the true continuum
is not seen redward of 2.29 $\mu$m because of the long wing of CO absorption
that makes the bandhead, but this is not important for our fitting
technique.

We observed multiple types of stars along with the galaxies; G and K giants
were observed as examples of templates for the velocity fitting 
and A dwarfs were
observed to obtain the shape of a ``flat'' spectrum.  These
dwarf stars have nearly featureless spectra in this 
region \citep{wal97}
and are extremely important to the data reduction.  The imager in
CoolSpec is cooled separately from the dispersive optics, requiring the
use of a filter just in front of the detector to reduce the thermal
background incident on the array. This filter shape must be removed
during data reduction. Also, the atmospheric absorption at
McDonald Observatory appears to vary on a timescale inconvenient
for these observations.  Atmospheric transmission calculations \citep{gaf95}
indicate that telluric absorption is dominated by CH$_{4}$, not H$_{2}$O, 
blueward of 2.34 $\mu$m, implying that atmospheric absorption should
not vary much with time.  This is not consistent with our experience at
McDonald Observatory, however; the shape of a single A star spectrum 
can change significantly during a night.  For both of these reasons,
it is very important to take a careful (and frequent) measure of the
detailed spectral shape of the filter/sky to be able to remove
this shape from the observed spectra.  We choose A dwarfs 
spatially near each
galaxy and observe one before and after
each galaxy/template observation.  We also observe an A dwarf sometime 
in the middle of long galaxy observations.

The observations are made by dithering the telescope 
30$^{\prime\prime}$ across
the slit to measure the sky at the same slit position in 
alternating exposures. Individual exposures are 120 seconds for
the galaxies and ten seconds for the stars.  Total integration times
for the galaxies vary from about one hour to almost five hours. Galaxies
that require very long integration times or that are at low declinations
are observed during several
nights to maintain reasonable airmasses.
The slit is rotated to the position angle of the galaxy major axis
as quoted in the RC3. 
Ar and Ne (or Xe for some runs)
emission lamps calibrate the wavelengths of the exposures.
Calibration exposures are taken every 24 minutes; 
the wavelength solution drifts significantly with time.
The telescope guides on either the galaxy itself
or a nearby star (if available) using the optical dichroic mirror autoguider.
No attempt is made to flux calibrate the spectra since we are mainly
concerned with the kinematic analysis.

Data reduction proceeds in several steps.  First, the images are 
rectified spectrally using the arc lamp emission lines.
We find that there is an additive
constant across the entire chip which varies from exposure to exposure,
so this dark current is measured for each exposure and then 
subtracted out.  We measure this in the same location for each
exposure, in pixels without signal from either slit position. 
We make a master background image
for each galaxy by masking out the object spectrum in each individual
exposure-- half of which have the object at one position on the slit, and
half of which have it at another position-- calculating the 
biweight, a robust estimator of the mean \citep{bee90}, of all the
exposures.
Galaxies observed on different nights, for long
periods of time, or during particularly humid nights required the 
construction of several different background
images to get good subtraction.  We find that we must choose carefully 
how much exposure time to 
average together: long enough to make a smooth, robust background image but
short enough to allow for changes in the sky.  This background image
is then subtracted from each individual exposure.  The images are next
shifted to the same wavelength solution; we interpolate between the two 
closest calibrations to find a good estimate of the wavelength solution and
then shift all the images from one observing run to the same solution.
This step must be done after the background subtraction or the image of
the seams between the detector's four quadrants does not subtract out well.\
All the images are then shifted so that the center of the galaxy in each
image is aligned; we calculate the biweight of all the processed
images to make one image for the galaxy.
The one-dimensional spectra are extracted from the 
two-dimensional images for basically the
entire galaxy; we choose the number of columns to extract to maximize
signal-to-noise.  For the sample, this varies between 
4$^{\prime\prime}$ and 20$^{\prime\prime}$. 
For some high S/N galaxies, we are also able to extract spatially
resolved spectra. The stellar spectra are reduced in a similar manner.  

To remove telluric absorption and the filter shape, the galaxy
and template stellar spectra must be divided by a ``flat'' spectrum.
After trying several approaches, we obtained the best results by using
the following procedure. First, all the A dwarfs from a run are averaged
together to make a smooth, high S/N sky spectrum. Figure \ref{Astarfig} 
shows such an average spectrum for the December 2001 run.  
The galaxy, template, and individual ``flat''
spectra are divided by this smooth spectrum.  For some galaxies and
template stars, this provides good flattening.  For the rest, the
spectrum is divided by a version of the A dwarf nearest in time to that
observation, smoothed
by the resolution element to reduce noise.  Dividing by only an individual
A star does not give results as good as the smooth sky spectrum made from
many A stars because of the
fluctuations in the individual spectra.  These individual stars change
because of fluctuations in the sky, not because of problems with S/N.


\psfig{file=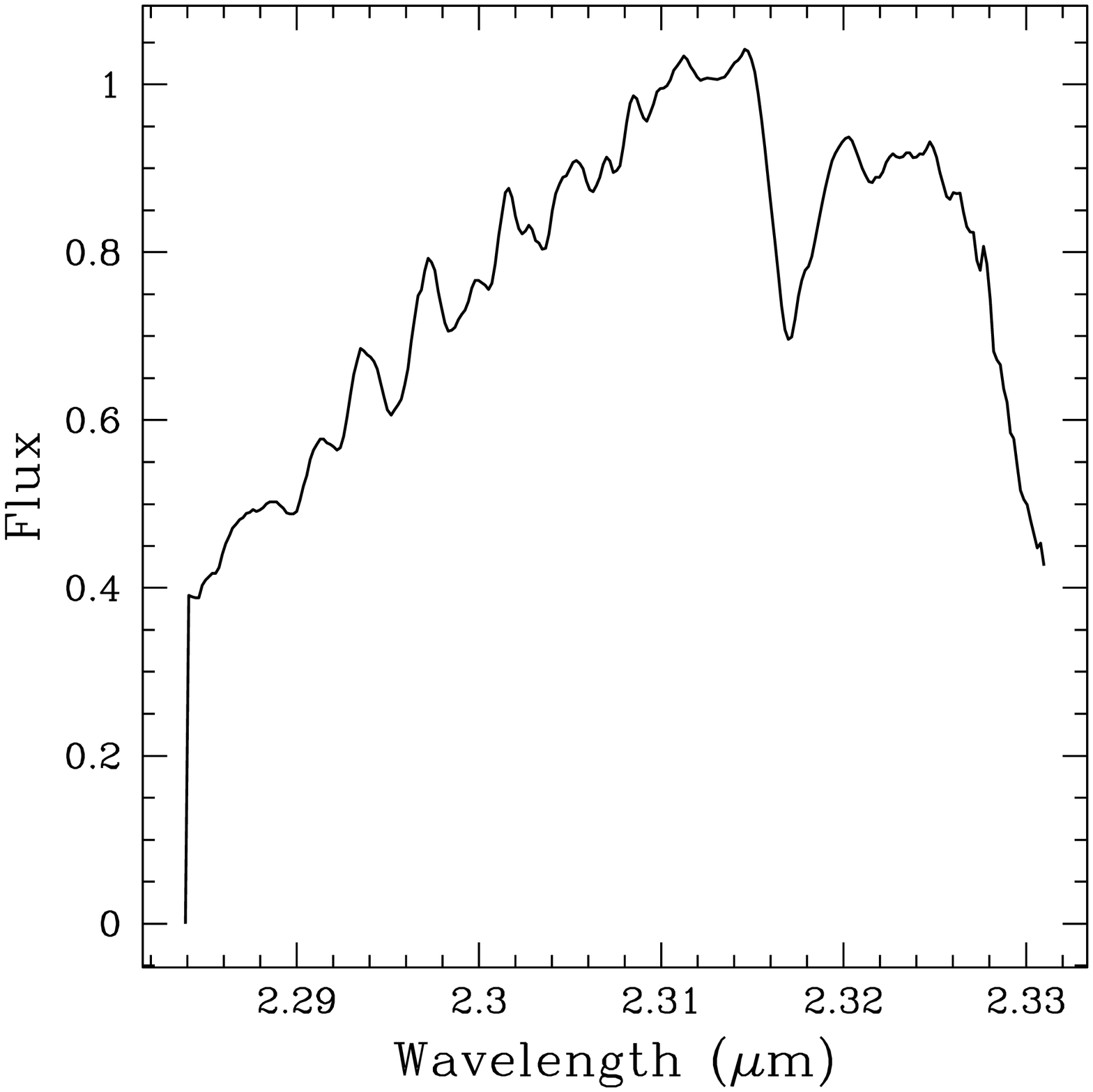,width=8.5cm,angle=0}
\figcaption[Astar.ps]{Average ``flat'' spectrum for 
December 2001 run.  This spectrum
is constructed from over 60 observations of A stars over the course of
the run, as described in the text.  The features in this spectrum 
are real and not due to noise.
\label{Astarfig}}


\section{Analysis}

\subsection{Extracting the Velocity Distribution}

A galaxy spectrum is the convolution
of the line-of-sight velocity distribution (LOSVD) 
with an average stellar spectrum.  
Figure \ref{conv} shows several examples; the dotted spectrum with
many small features is that of a typical K giant while the smooth spectra
are the stellar spectrum convolved with three Gaussian LOSVDs (with 
dispersions of
100, 200, and 300 km s$^{-1}$). This figure shows the smoothing and broadening
due to the internal velocities. The bandhead is obviously very broad,
but its sharp blue edge allows us to measure accurately the
kinematics.

There are several techniques used to obtain the internal kinematic 
information from a galaxy spectrum.
The cross-correlation technique \citep{ton79} extracts the
velocity dispersion utilizing
the cross-correlation function of the galaxy
spectrum and template stellar spectrum; the width of the peak of the
cross-correlation function provides the dispersion.  This technique
requires correct subtraction of the continuum and is sensitive to
template mismatch.  The Fourier quotient technique \citep{sar77} 
uses
the result of the convolution theorem: the Fourier transform of the
LOSVD is the quotient of the Fourier transform of the galaxy spectrum 
and the Fourier transform of the template spectrum.  
Both of these techniques fit the kinematics in Fourier space;
however, the spectrum can
also be fit directly in pixel space, which is the approach taken here.

We use the fitting technique of
\citet{geb00}, deconvolving the spectrum using a maximum penalized 
likelihood estimate to obtain a nonparametric LOSVD. An initial velocity
profile is chosen and this profile is convolved with a stellar template
spectrum.  The residuals to the galaxy spectrum are calculated and the
velocity profile is changed to minimize the residuals and provide the
closest match to the observed galaxy spectrum.


\psfig{file=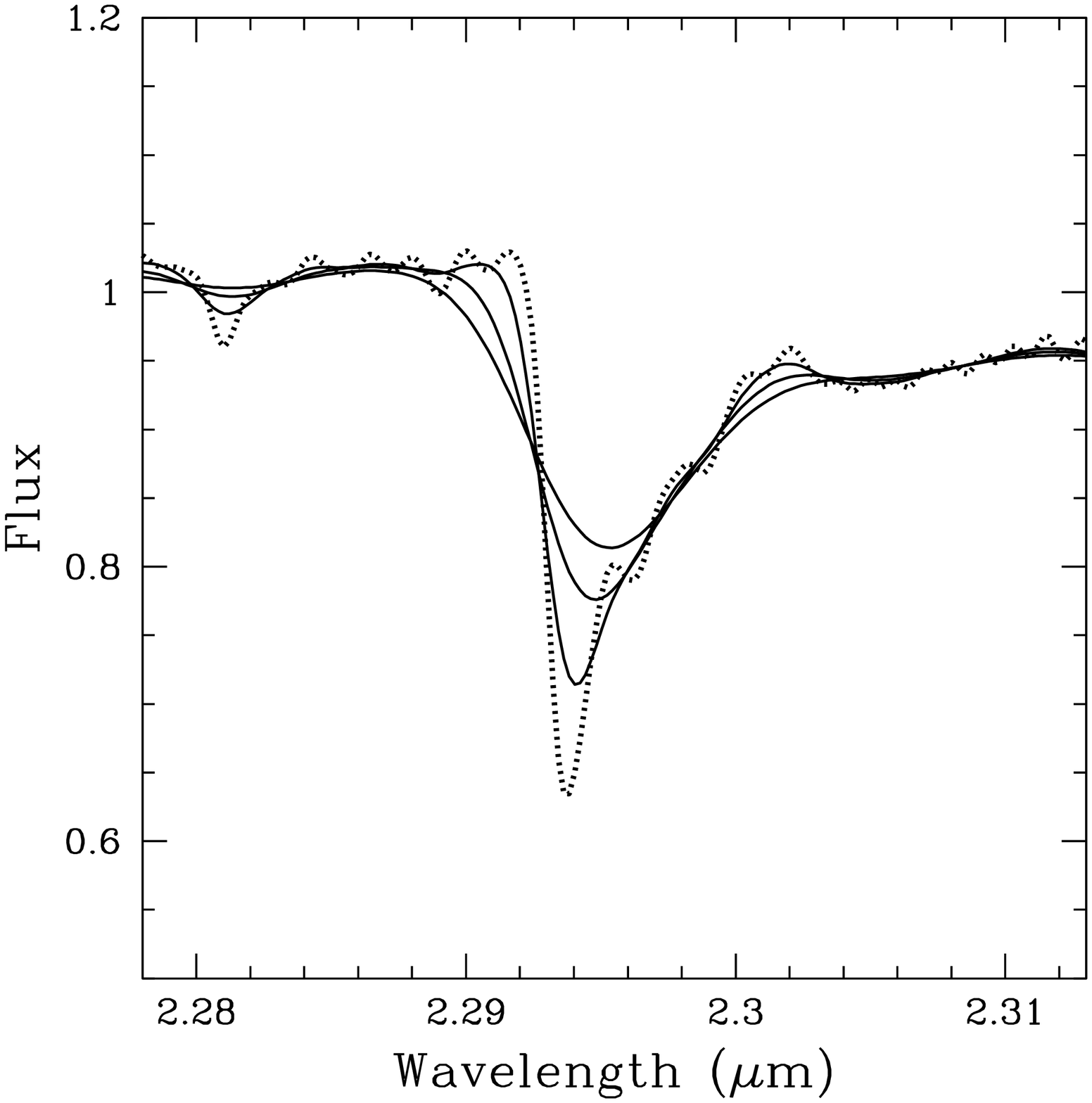,width=8.5cm,angle=0}
\figcaption[convolving.eps]{CO bandhead in a 
typical mid-K giant (dotted line with
high frequency features) at our resolution and then
convolved with Gaussian
LOSVDs with dispersions of 100, 200, and 300 km s$^{-1}$ 
(solid spectra).  The equivalent width for this star is 13.4 \AA.
\label{conv}}
\vskip 10pt


The choice of template star proves to be important for the fitting
results.  Previous work in the CO bandhead 
used mostly K and M giant stars as templates, but it is not clear
that this is a correct choice.
We use the atlas of \citet{wal97} to test
the effect of stellar type on the fitting results.  The equivalent
width of the CO bandhead is a function of the effective temperature
and surface gravity of the star; either increasing the surface gravity 
or decreasing the temperature increases
the equivalent width.  Figure \ref{EWtype} shows
our calculations of the equivalent widths of most of the stars of
\citet{wal97}; these trends are evident.


\psfig{file=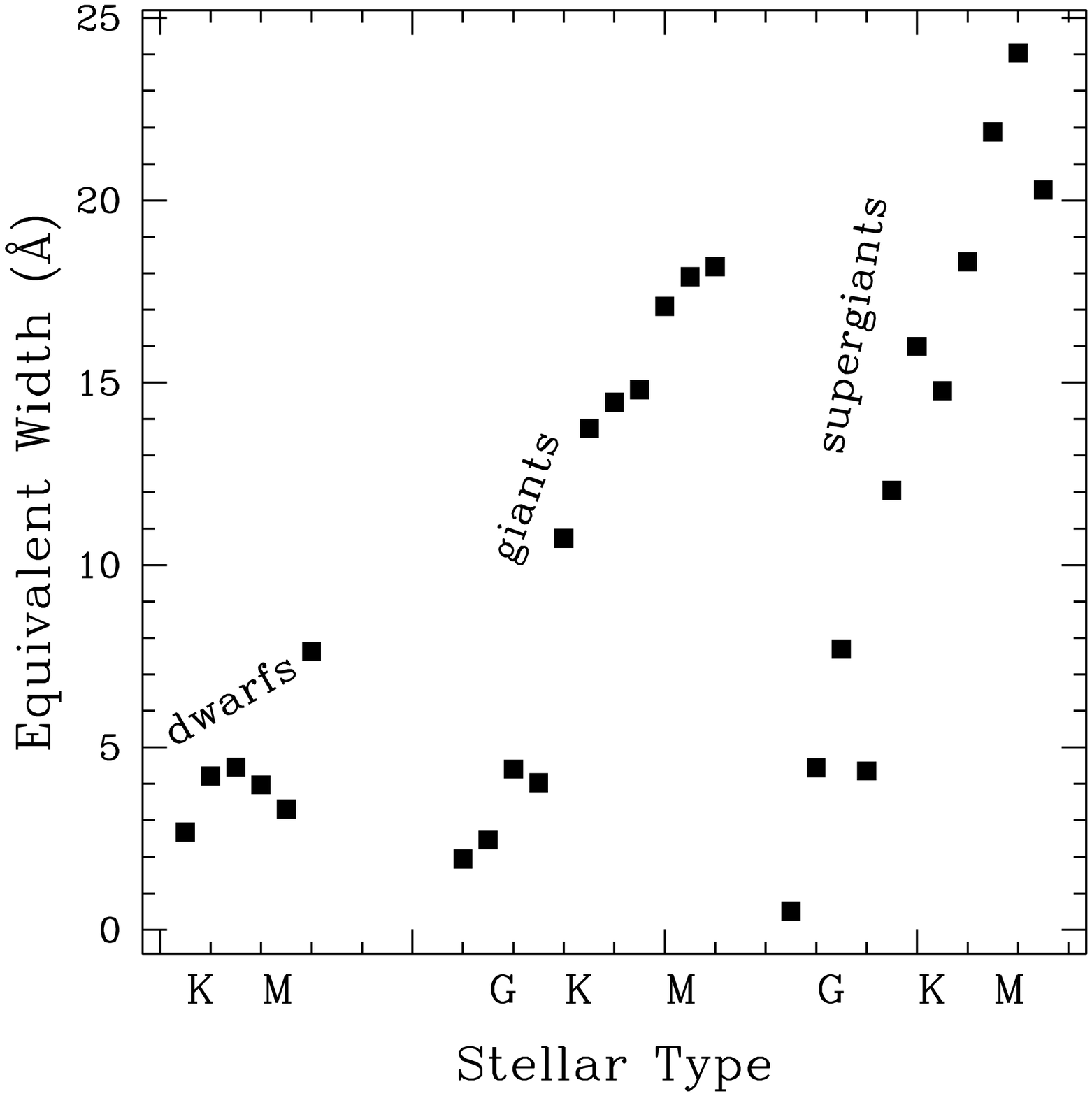,width=8.5cm,angle=0}
\figcaption[EWtype.eps]{Equivalent width of the CO bandhead as
a function of spectral type.  These are our measurements
of equivalent width for the stellar spectra of \citet{wal97}.
\label{EWtype}}
\vskip 5pt


\pagebreak
We have found that
the dispersion measured by the fitting program 
depends on the template spectrum chosen for
the fitting.  Figure \ref{EWdisp} illustrates this point for 
NGC 1161.  The dispersion measured for this galaxy increases as the
equivalent width of the template star's bandhead increases.  
The $\chi^{2}$ for each of these fits is shown as well and is
is quite high for most, indicating 
poor fits. One template with an equivalent width near
the galaxy's equivalent width appears to give a significantly better fit.
Other galaxies show a similar trend.  The shape of the CO
bandhead must be a function of equivalent width.  To obtain a reliable
dispersion measurement, we give the fitting program a variety
of template stellar spectra and allow it to vary the weights given
to the different stars to obtain the best fit.  As a result, along with
the LOSVD information, the fitting program also provides stellar
population information.  We have explored the effect on the fitting 
of using stars of the same equivalent width but different stellar types;
this does not seem to be important.  It is the equivalent width of the
template that counts, not the details of the spectral type.  We choose
eight stars from \citet{wal97} to use as our available
templates.  These stars have a range of equivalent
widths, ranging from less than 5 \AA~to over 20 \AA.  The best fit
almost always gives most of the weight to a few
of the template stars.  We have also
used template stars observed with the same instrumental
setup used for the galaxies.  We find similar results using
these stars but the better S/N and larger equivalent width
variation of \citet{wal97} make that dataset more useful.
These spectra have a somewhat higher spectral resolution than
ours, so before using them as stellar templates we have carefully
convolved them to our spectral resolution.


\psfig{file=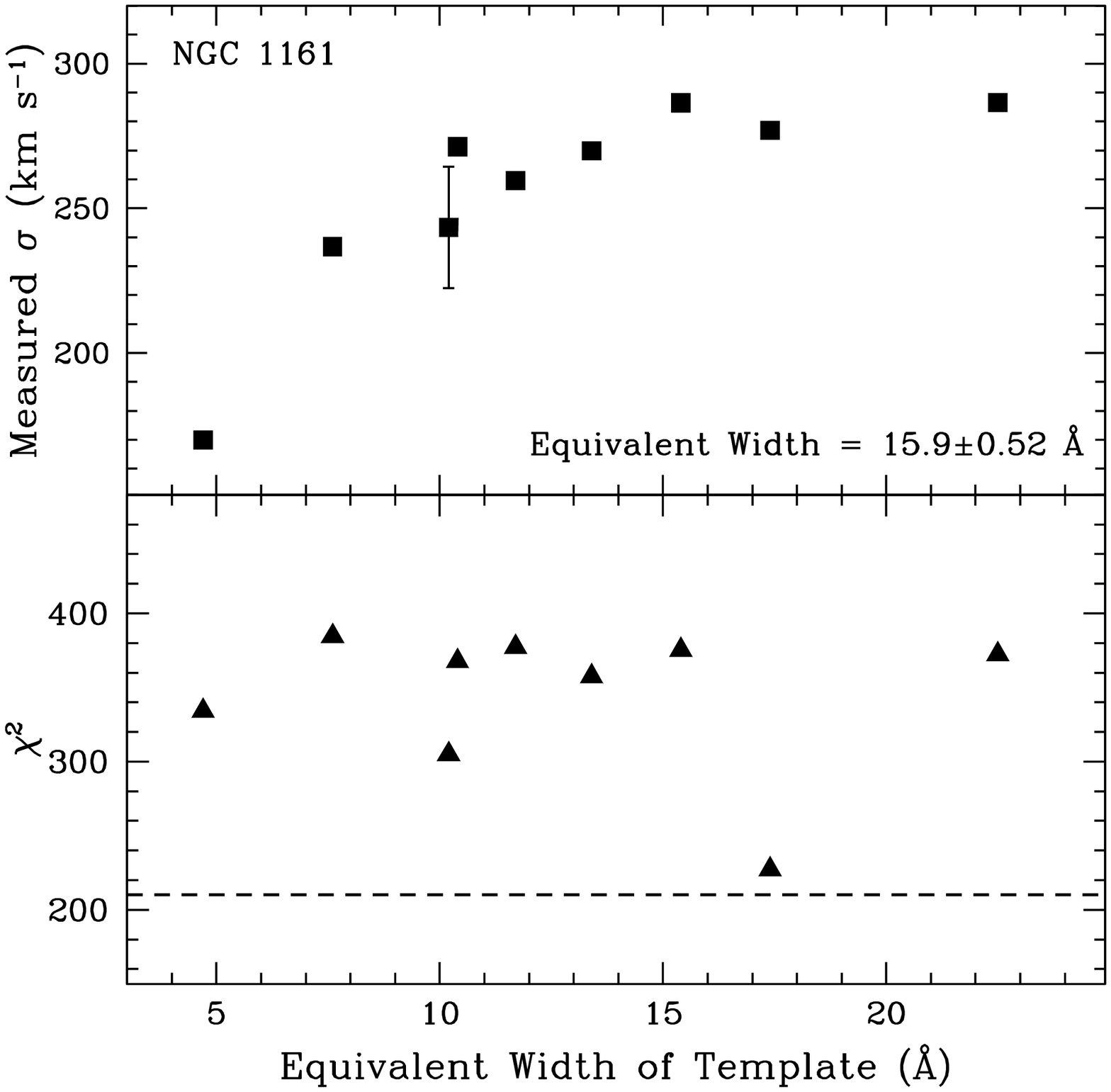,width=8.5cm,angle=0}
\figcaption[EWdisp.eps]{Dispersion measured 
by the fitting program for NGC 1161
as a function of the equivalent width of the input template star
and the $\chi^{2}$ for each fit.
The error bar shown is representative for all these dispersions.
For comparison, the dispersion measured for NGC 1161 allowing the
program to choose templates is 274 km s$^{-1}$ with 
$\chi^{2}$ = 211, lower than any of the fits using individual stars.
The dashed line shows the $\chi^{2}$ for an acceptable fit, given
the 210 constraints.
\label{EWdisp}}
\vskip 10pt


Because of this equivalent width effect, the continuum choice
may skew the velocity dispersion measurement.  If the red continuum
chosen is too blue, i.e. encroaching on the bandhead, the
measured equivalent width will be too low and the resulting
measured velocity dispersion will be lower than the true
dispersion.  We have tested the effects of the continuum choice
and find that this is not a large source of uncertainty. We convolved
a stellar template with Gaussian LOSVDs and made different choices
for the red continuum.  The effect was as expected but was not large.
Only very obviously bad choices for the red continnum cause a 10\%
decrease in the dispersion; more realistic mistakes in the continuum
choice cause decreases of less than 5\%, smaller than most of our
error bars.  We have also examined the effects of changing
the continuum definition in the actual galaxy spectra; the changes in the
LOSVD fits and velocity dispersions are very small.

\begin{figure*}[t]
\centerline{\psfig{file=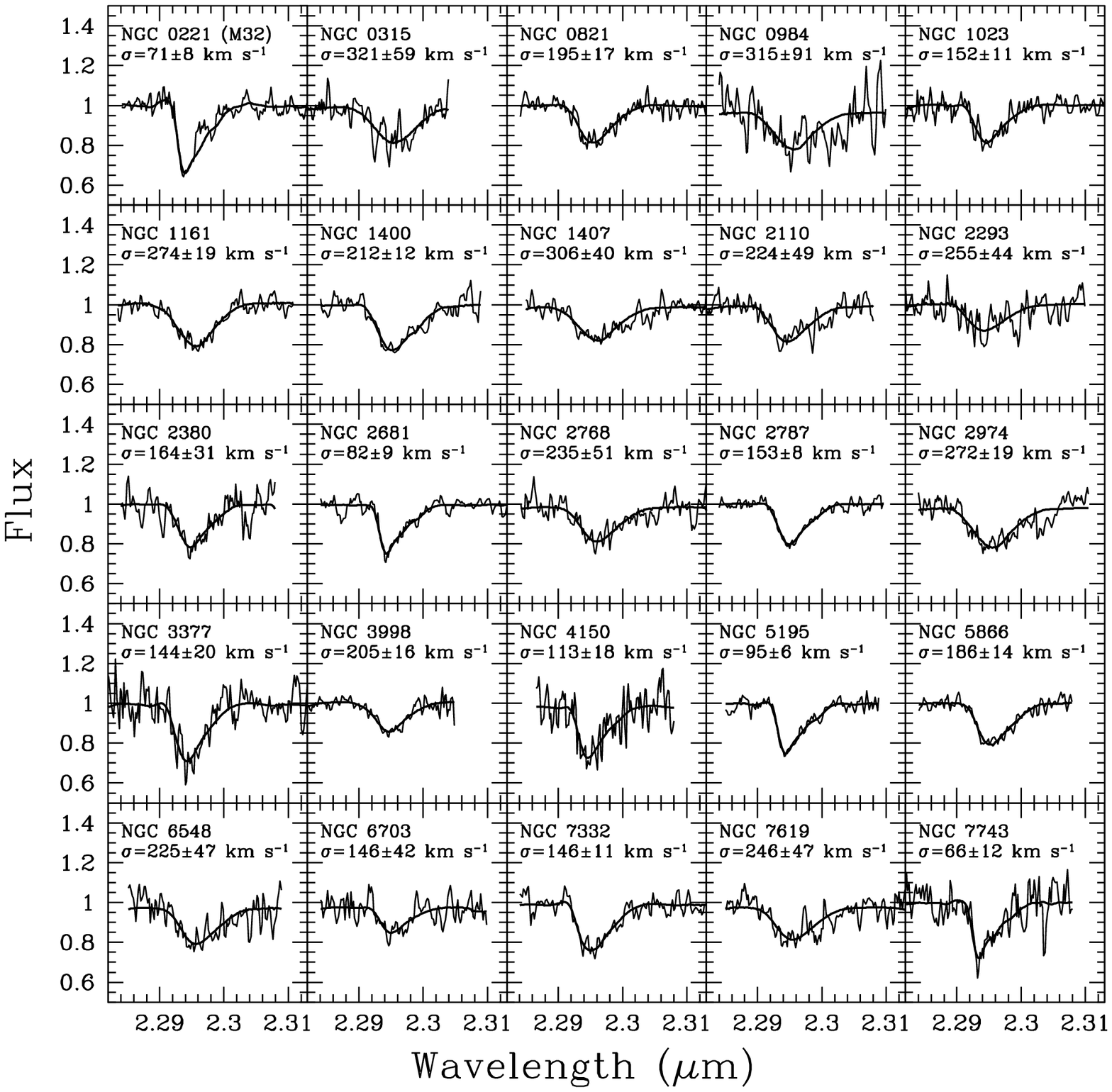,width=20cm,angle=0}}
\figcaption[galspecall.eps]{Rest-frame spectra for 
galaxies observed at McDonald Observatory (noisy line) and for the
template stellar spectrum 
convolved with the derived velocity distribution (smooth
line).  The derived velocity dispersion and its 68\% uncertainty
are reported for each galaxy.
\label{gal1}}
\end{figure*}

This fitting technique obtains a nonparametric LOSVD; no a priori
assumptions about the shape of the LOSVD are made (except that it
is nonnegative in all bins).  To measure a dispersion from this
nonparametric LOSVD, we fit a Gauss-Hermite polynomial to it and
use the second moment as the dispersion.
We can also fit a Gaussian LOSVD directly
to the spectrum.  Galaxies in the sample with low S/N required the
assumption of a Gaussian LOSVD in order to achieve a sensible
velocity distribution.  The fitting program can get lost
in residual space when there are too many free parameters
for the level of noise in the spectrum,
and this was the case for the lower S/N galaxies in our sample. We
compared the derived nonparametric and Gaussian LOSVDs for galaxies
with higher S/N and found good agreement between them.
Figure \ref{gal1} shows the results for 
all 25 sample galaxies and 
the derived velocity dispersion for each galaxy. The noisy line
is the observed spectrum for each galaxy and the smooth line is
the template stellar spectrum convolved with the derived LOSVD.

The uncertainties for these galaxies are determined using the 
Monte Carlo bootstrap
approach of \citet{geb00}.  For each galaxy, a template spectrum 
is convolved with
the derived LOSVD to make an initial galaxy spectrum.  We then
generate 100 realizations of the galaxy spectrum by randomly drawing
the flux values at each wavelength from a Gaussian distribution.
The mean of this distribution is the value of the initial template
spectrum convolved with the LOSVD (i.e. the smooth line in 
Fig. \ref{gal1}) and the standard deviation is the 
root-mean-square of the initial fit.
These 100 synthetic galaxy spectra are then deconvolved to
determine the LOSVDs.  These LOSVDs provide a distribution of
values for each velocity bin which allow us to estimate the uncertainty
and examine any bias in the dispersion. To generate the 68\%
confidence bands, we choose the 16\% to 84\% values from the 100
realizations.  The median of the distribution determines any
potential bias from the initial fit. Figure \ref{SNvsaccuracy} shows
how the S/N of the observed galaxy spectrum affects the percent
accuracy to which we can measure the dispersion.  Using this technique
and spectral feature, 10\% accuracy in the velocity dispersion
requires S/N per pixel of about 25 or 30.


\pagebreak

\clearpage
\psfig{file=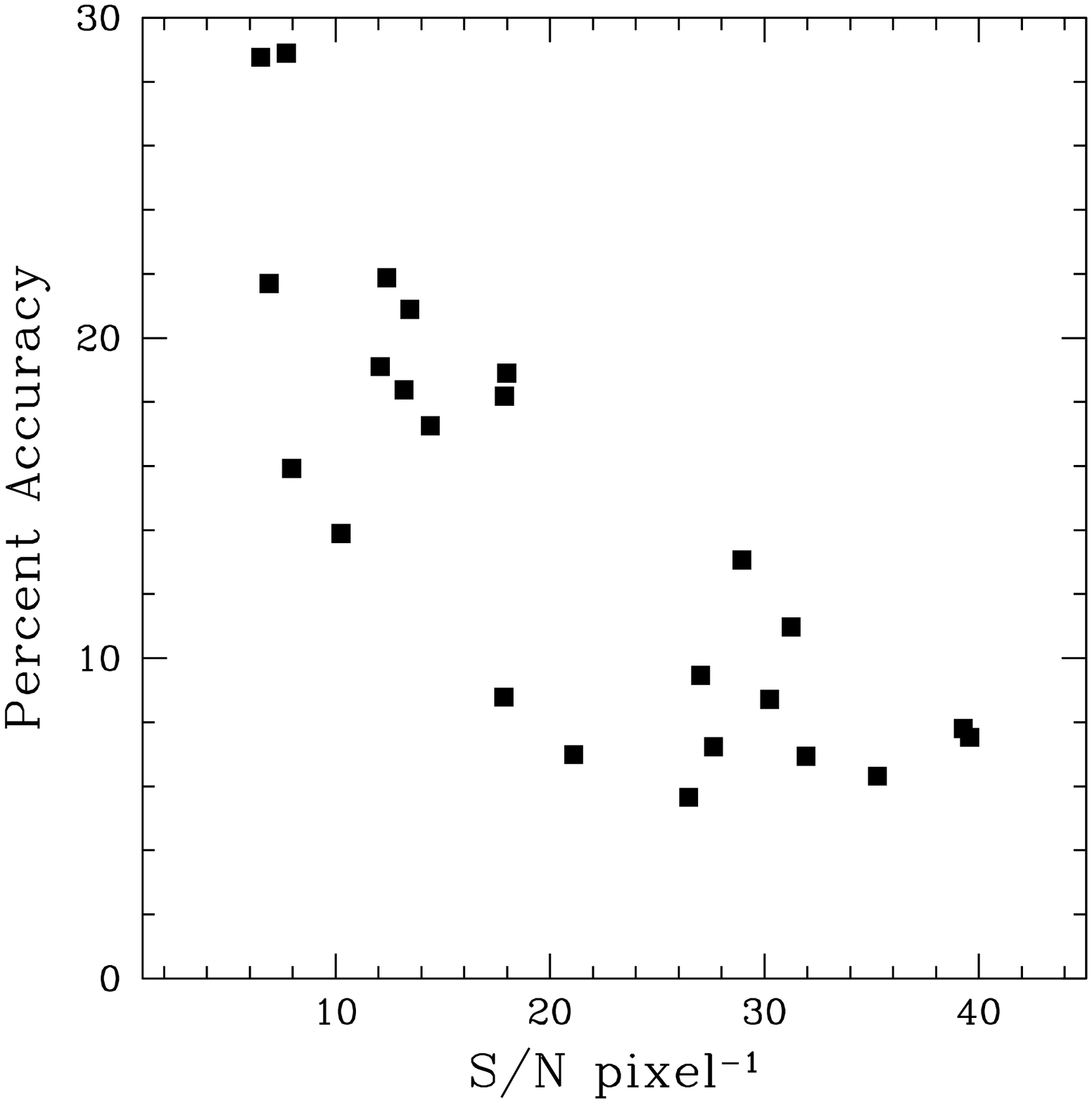,width=8.5cm,angle=0}
\figcaption[accuracy.eps]{Percent accuracy to 
which we measure the velocity dispersion
using the fitting technique described in the text versus measured
S/N per pixel of the spectra. Our wavelength scale is 1.9 \AA/pixel.
\label{SNvsaccuracy}}


\psfig{file=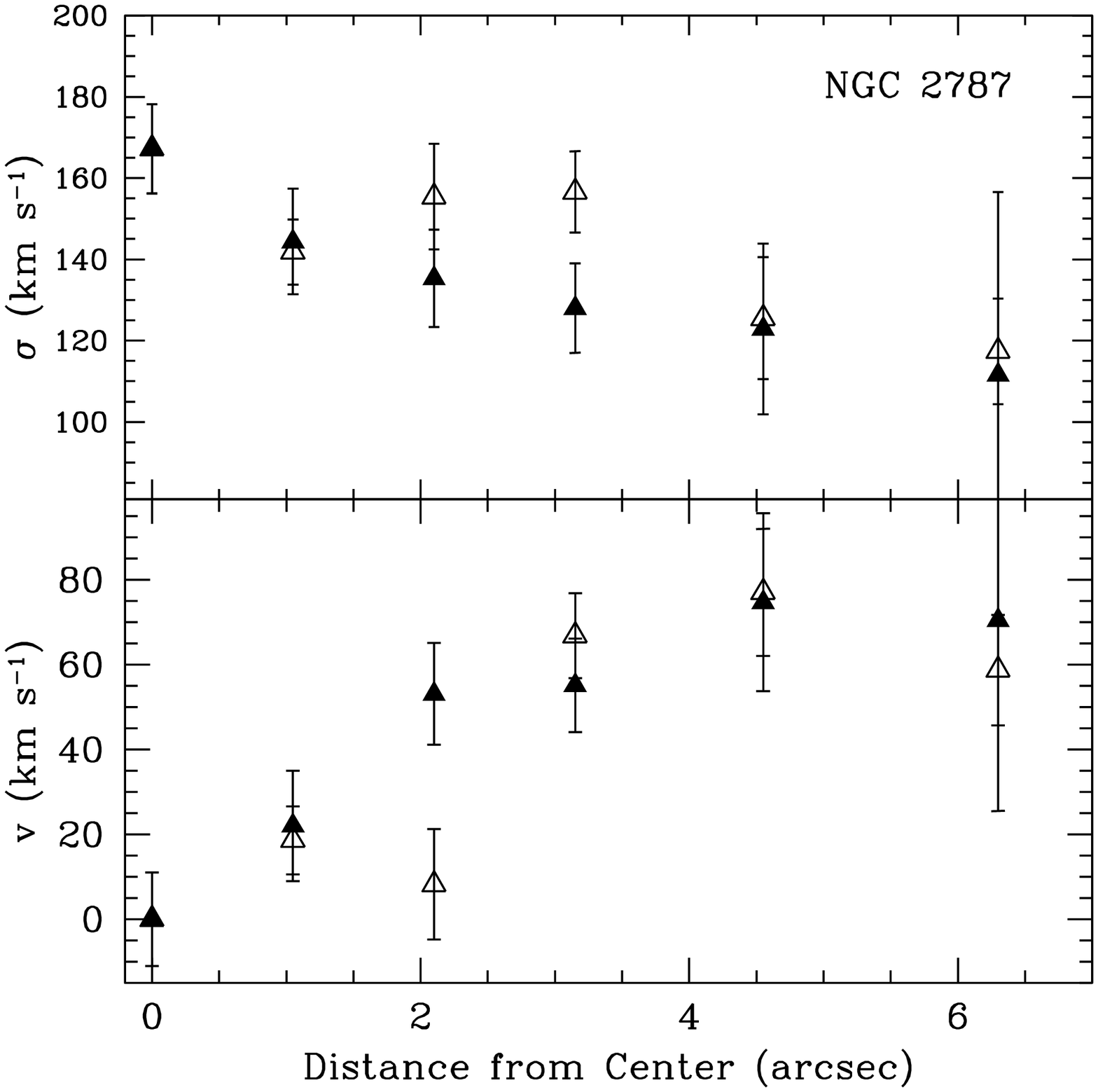,width=9cm,angle=0}
\figcaption[N2787spatial.eps]{Velocity dispersion and 
velocity curves for NGC 2787. The open and filled triangles
are for the opposite sides of the galaxy. The velocity dispersion derived
for the entire galaxy as described in the text is $153 \pm 8$ km s$^{-1}$. 
\label{N2787fig}}


It is possible to obtain spatially resolved kinematics from these data
as well, when the S/N for the galaxy is high enough.  
Figure \ref{N2787fig} shows
the variation of velocity and
velocity dispersion with distance from the center of
the galaxy for NGC 2787.  In the center regions, we extract spectra of
approximately the width of the seeing disk for that night and in the 
outer regions, we increase the number of pixels extracted to keep
the S/N high enough to derive the LOSVD.  The velocity dispersion
curve is peaked in the center of the galaxy, and both sides of the
galaxy show the same decrease.  The velocity dispersion measured for
the entire galaxy (extracted to maximize S/N) is $153 \pm 8$ km s$^{-1}$.
Figure \ref{N2787fig} shows that this is the value achieved
by the velocity dispersion curve between 0.5 and
1.5 arcseconds away from the center.

\subsection{Equivalent Widths}

The equivalent width of the 2.29 $\mu$m CO feature can be used
to quantify stellar population effects in these
galaxies.  To measure the equivalent width, we define
continuum on both sides of the feature and fit a straight line
between these two points.  In the rest frame of the galaxy,
the continuum on the blue is defined to be the median between 2.287
and 2.290 $\mu$m, and on the red between 2.304 and 2.307 $\mu$m 
(about 15 pixels on each side).  We use the velocity information
from the LOSVD fitting procedure to shift the galaxy spectra to their
rest frames.
We measure the area between
the continuum line and the observed spectrum between 2.290 and 2.304
$\mu$m, and divide by the
continuum to find the equivalent width.  Since true
continuum is not seen redward of the feature, we choose
to define the quasi-continuum where the spectrum 
again becomes nearly horizontal.

We calculate uncertainties for the equivalent width using the same 
Monte Carlo method used for the velocity dispersion. We generate
100 realizations of each galaxy spectrum from the fit to the
observed spectrum found by the LOSVD fitting procedure, with noise
chosen to match the S/N of the observed spectrum.  We measure
the equivalent width of these 100 synthetic galaxy spectra and
use this distribution of equivalent widths to estimate the
uncertainty and examine bias. We choose the 16\% to 84\% values
from this distribution for our 68\% uncertainty, and use
the median to look for bias.  We also measure the equivalent
width of the (noiseless) fit from the LOSVD extraction, 
which matches the median from the Monte Carlo simulations well.

There does not appear to be any bias for most of these galaxies,
i.e. the equivalent width of the observed spectrum is close to
(within the uncertainty of)
the median from the Monte Carlo simulations.  For three galaxies,
this is not the case.  Messier 32, NGC 984, and NGC 2974 all have
features in their spectra that cause a significant difference between
the equivalent widths of the observed spectra and the fits
(see Fig. \ref{gal1} to see these differences).  
These features are likely due to variability
in sky absorption which was not well-removed by our flattening
procedure, combined with low S/N for NGC 984 and NGC 2974.  
(See later section
on Messier 32 for more discussion.)  For these galaxies, we choose to
use the equivalent width of the fit, basically the same
as the median from the Monte Carlo simulations, as our value
for the equivalent width.  We believe these values are more reliable
than the values compromised by sky variability and noise issues.

These measurements of the equivalent width must be corrected for the
effect of the galaxy dispersion.  The velocity broadening throws
equivalent width outside of our chosen measuring region, especially
on the sharp blue edge. To calculate this correction, we take spectra
for several types of stars, with equivalent widths ranging from about 7
\AA~to almost 20 \AA, and convolve them with Gaussian
velocity distributions of different dispersions.  We then measure
the equivalent widths of these convolved spectra in the same way we measure
the galactic spectra in order to see
how much we should correct our real equivalent width
measurements.  The dispersion correction for the CO bandhead
agrees well between the different
stars used, and we find this correction to be rather large, about
20\% for 300 km s$^{-1}$.  This is typical for some features
used in optical regions \citep{tra98}.
The correction we use, found by fitting a cubic
polynomial to the data from the convolved stellar spectra, is
$$\frac{EW_{observed}}{EW_{actual}} = 1-1.853\times 10^{-4}\sigma+1.287\times 10^{-6}\sigma^{2}-9.695\times 10^{-9}\sigma^{3}$$
where $EW_{observed}/EW_{actual}$ is the ratio of observed
and actual equivalent widths and $\sigma$
is the velocity dispersion measured in km s$^{-1}$. Figure \ref{EWcorrect}
illustrates this dispersion correction.  The symbols show the fractional
change in equivalent width for each star as a function of velocity
dispersion, and the line shows the fit mentioned above used to correct
our galaxy equivalent width measurements.


\psfig{file=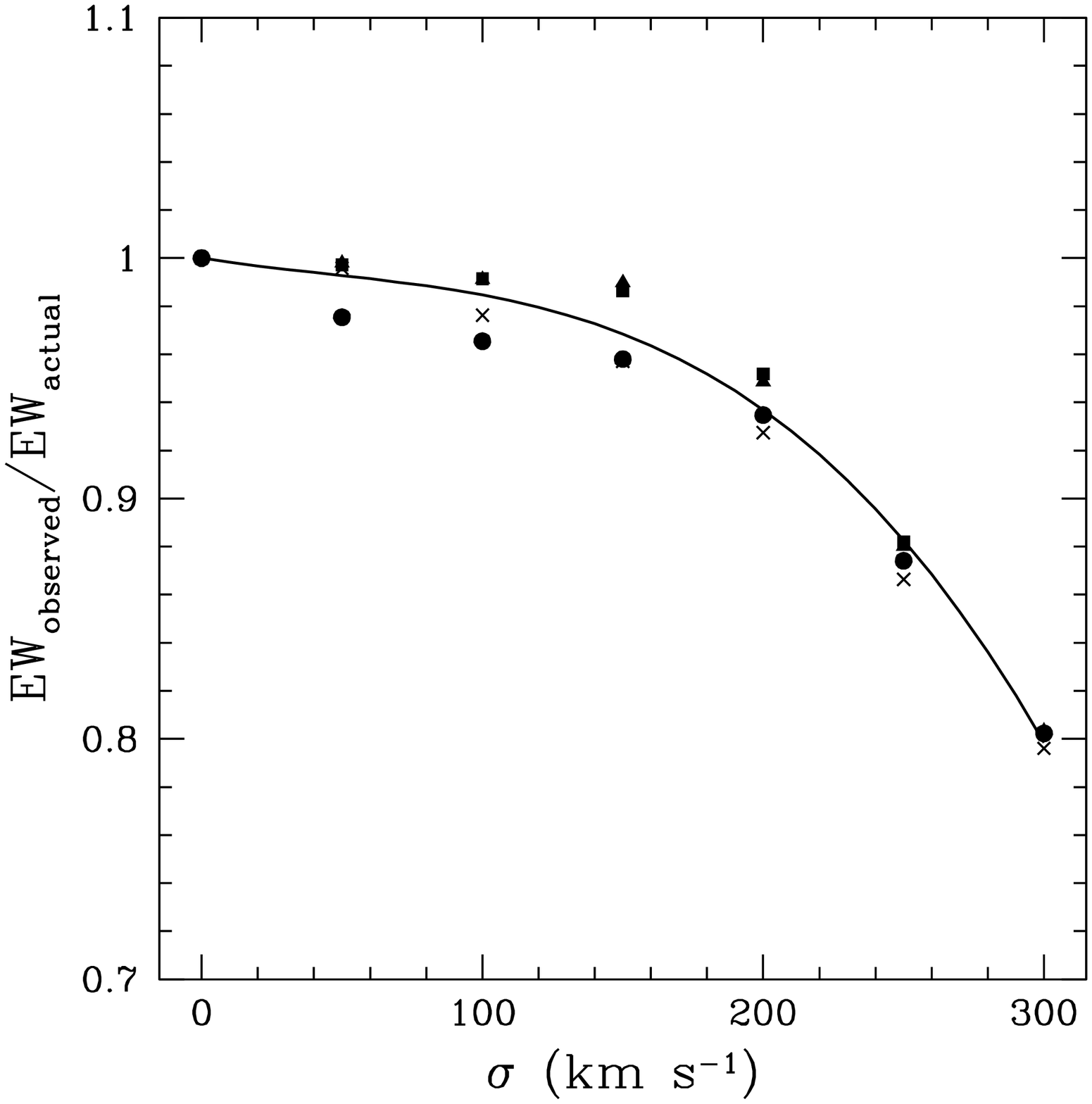,width=8.5cm,angle=0}
\figcaption[EWcorrection.ps]{Dispersion correction to the equivalent width
of the CO bandhead.  The different symbols represent the
change in measured equivalent width for stars 
with a range of equivalent widths convolved
with Gaussian LOSVDs; the line is a fit to these data (described in
the text) used to correct our galaxy equivalent width measurements.
\label{EWcorrect}}


\subsection{Dust Masses}

In order to ascertain the effects of dust on observed kinematics,
we need to know how much dust there is in individual galaxies.
In this section, we derive dust masses for our sample galaxies
from IRAS flux densities according to the technique of \citet{gou95}.
The notation and equations below are identical to those presented
in \citet{gou95}.
The 60 and 100 $\mu$m flux densities are sensitive mostly to emission
from cool (i.e. interstellar) dust, but are contaminated by hot 
(i.e. circumstellar) dust.  \citet{gou95} calculated this contribution
using data from Galactic Mira stars and found corrections to be
$$S(60)_{corr} = S(60)-0.020S(12)$$
$$S(100)_{corr} = S(100)-0.005S(12)$$
where $S(12)$ is the 12 $\mu$m IRAS flux density.  These corrections
are small for the sample galaxies.

The mass of dust $M_{d}$ in a galaxy is
$$M_{d} = \frac{D^{2}S_{\nu}}{\kappa_{\nu}B_{\nu}(T_{d})}$$
where $D$ is the distance to the galaxy, $S_{\nu}$ is the flux
density, $B_{\nu}(T_{d})$ is the Planck function for the dust
temperature $T_{d}$ at frequency $\nu$, and $\kappa_{\nu}$ is the
dust opacity
$$\kappa_{\nu} = \frac{4\pi a^2}{3\pi a^3 \rho_{d}}Q_{\nu}$$
where $\rho_{d}$ is the specific dust grain mass density, $a$ is 
the average grain radius weighted by grain volume, and $Q_{\nu}$
is the grain emissivity factor.  Using the grain size distribution
from \citet{mat77} and the values of $Q_{\nu}$ from \citet{hil83},
the dust mass in solar masses is
$$M_{d} = 5.1\times 10^{-11}S_{\nu}D^{2}\lambda^4(\exp(1.44\times 10^4/\lambda T_{d})-1) M_{\odot}$$
where $\lambda$ is in $\mu$m, $D$ is in Mpc, and $S_{\nu}$ is in mJy.

We calculate the dust temperature as the color temperature determined
from the corrected $S(100)/S(60)$ flux ratio using the assumption of
a dust grain emissivity law $\propto \lambda^{-1}$,
typical of astronomical silicates at wavlengths $\lambda \la
200 \mu$m \citep{hil83,row86,mat89}:
$$\frac{S(60)}{S(100)} = \left(\frac{\nu_{60}}{\nu_{100}}\right)^4 \frac{\exp(h\nu_{100}/ kT_{d})-1}{\exp(h\nu_{60}/ kT_{d})-1}$$
where $h$ is the Planck constant, $k$ is the Boltzmann constant, 
and $\nu$ is the frequency in Hz.
The temperatures calculated in this way should be regarded as
estimates because a temperature distribution is surely more realistic
than isothermal dust.  
Also, IRAS is sensitive to cool dust with $T_{d} \ga$ 25 K, but it 
is predicted that much of the
dust in a normal galaxy will be colder ($\simeq$ 10-20 K) and will 
emit more strongly at longer wavelengths. IRAS provides little 
information on this cold dust. Thus, the masses calculated from the 
IRAS color temperature
are lower limits on the total dust mass in a galaxy and may be over an order
of magnitude too low \citep{kwa92,gou95,wik95}.

Table \ref{tbl-2} lists the IRAS flux densities from  \citet{kna89}
and the 1994 correction to those data, calculated dust temperature, 
and calculated dust
mass for each galaxy in this sample. For galaxies which were not detected at
both 60 and 100 $\mu$m, the dust temperature is assumed to be 30 K.  
For galaxies that are not detected in either band, the 
dust mass shown is an upper limit.
As expected, the calculated dust temperatures lie between 25 and 50 K, the
range of IRAS's greatest sensitivity.  The galaxies in this sample have a wide
range of dust masses, from no detectable mass to over a million solar masses.
This is, of course, a tiny fraction of the total mass of a galaxy and does
not affect the potential, but may significantly affect the observed
kinematics.

\section{Results and Discussion}

Table \ref{tbl-3} shows the dispersions measured from the CO bandhead and 
the optical dispersion measurements from the literature we
use as comparisons.  We are aware of the difficulties in making
such comparisons; optical stellar kinematic data can vary significantly
between authors, especially in the dispersion. 
For each galaxy, we thoroughly investigate the published 
kinematic data and carefully choose the measurement which is the most 
reliable. We make these choices based on S/N ratio, LOSVD-fitting 
technique, and the concordance between recent published dispersions.

The homogeneous optical sample with the most galaxies in common
with this study is that of \citet{tra98}, a large reliable
study of early-type galaxies.  \citet{tra98} has measurements
for 15 of our 25 galaxies.  We have worked through the analysis 
in the following sections using only these galaxies and their 
dispersions from this paper; we find no significant changes in 
the results reported below.  Our results do not appear to be 
dependent on author-to-author variations in published kinematics. 
The reader will notice in table \ref{tbl-3} that we did not always choose
to use the \citet{tra98} data when available.  In some cases there
are more recent measurements which appear to be better. These newer 
measurements are almost always consistent with \citet{tra98}.  
The exception is NGC 3377.  We use 145 $\pm$ 7 km s$^{-1}$ from 
\citet{kor98}, while \citet{tra98} quotes 126 $\pm$ 2 km s$^{-1}$.  
We believe this discrepancy is due to the high rotation of NGC 3377, 
which \citet{kor98} includes but \citet{tra98} appears not
to have included.

There are three galaxies in our sample which have particularly
unreliable optical measurements.  These are galaxies which have
especially large differences between reliable published optical
dispersions, or which have only very old dispersion measurements.  
These galaxies are NGC 4150, NGC 5195, and NGC 6548.  We repeated 
the analysis in the following sections without these galaxies 
in the sample, and again find no significant changes in our results.

For each galaxy, we choose an integrated optical dispersion 
measured with a similar extraction window to our measurement.
However, the integrated dispersion as a function of extraction
window does not vary significantly \citep{geb00b}.  Thus, any
errors made due to uncertainty in the extraction window
comparison will be minimal.  For example, NGC 3998, which has
steeply rising kinematic profiles, shows no variation in the
derived velocity dispersion when we use different extraction
windows. For optical 
dispersions without quoted uncertainties, we assume 5\% of the 
measured value.  Also, for galaxies with quoted uncertainties 
less than 5\%, we substitute 5\% of the measured value, assuming 
that problems with template mismatch, continuum subtraction, etc. 
introduce uncertainty of at least that magnitude into the 
optical uncertainties.

\subsection{Comparing IR and Optical Dispersions}

Figure \ref{dispfig} shows the correlation between the CO dispersions
and the optical dispersions.
The most obvious aspect of this comparison is that most of the CO
measurements are lower than the optical dispersions.
The dashed line has a slope of unity, showing where the two
measurements are equal, and the solid line is a fit by least squares
to the data.  The slope of the best fit line is 1.189 $\pm$ 0.084,
which is a significant difference from unity at a 2.25$\sigma$ level.
The intercept of the best fit line is -8.6 $\pm$ 12.4, not 
significantly different from zero.
The $\chi^{2}$ of this best fit is 54.4.  If we assume that there
should be a correlation between the dispersions, then the
24 constraints imply that we have underestimated our uncertainties.
To have a $\chi^{2}$ value of 24 (matching the 24 constraints),
all the errors need to be scaled up by 50\%. The $\chi^{2}$
of the dashed line in the plot (i.e. equality between optical and IR
dispersions) is 76.2, a markedly worse fit. Increasing our
uncertainties by 50\% is unrealistic and the large scatter is
likely real, possibly reflecting random dust distribution
between galaxies. 

The true ellipticals are represented
in Figure \ref{dispfig} as open triangles 
while the S0s are filled circles.  Our
sample does not contain many true Es (because of the make-up
of the SBF sample at these redshifts) so it is difficult to make
definitive statements, but we examine the differences between
the two populations.  The true ellipticals appear to be
more consistent between the infrared and optical data.
The slope of the best fit line to only the ellipticals
is 0.994 $\pm$ 0.091, consistent with unity, and the
intercept is 4.8 $\pm$ 13.4, consistent with zero.
The $\chi^{2}$ of this best fit is 1.77, which is at least
consistent with the seven constraints.


\psfig{file=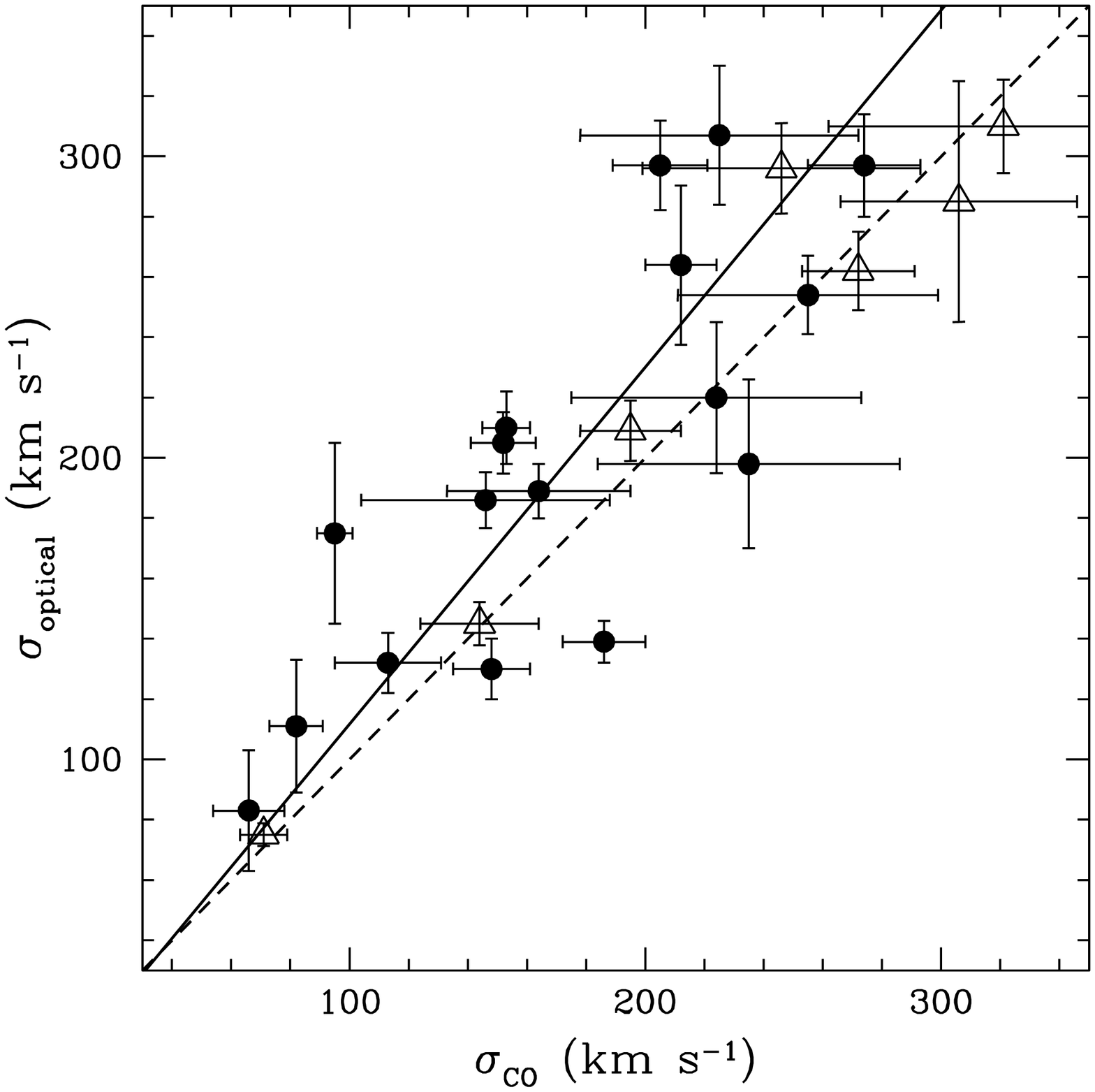,width=8.5cm,angle=0}
\figcaption[sigma.eps]{Correlation between the 
dispersion measured from the CO
bandhead and the optical dispersion from the literature. The dashed
line has a slope of unity, showing where the two measurements are
equal. The solid line is the best fit to the data, as described in
the text.  Here and in following plots, the filled circles are 
S0 galaxies while the open triangles are true Es.
\label{dispfig}}


\psfig{file=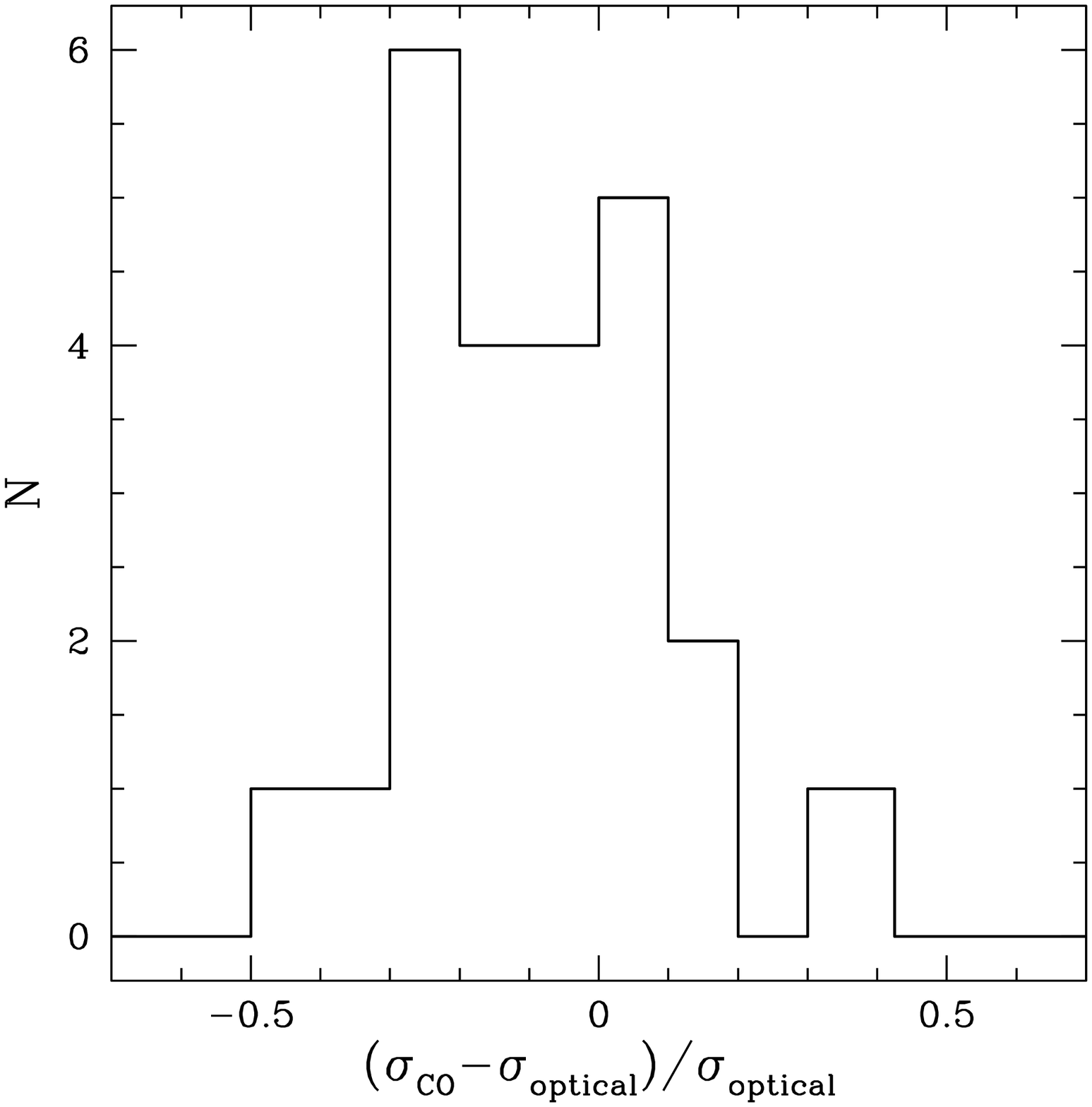,width=8cm,angle=0}
\figcaption[histogram.eps]{Histogram showing the 
number of galaxies in each bin of
fractional difference between infrared and optical measurements
of dispersion.  The median difference is 11\% smaller.
\label{histogramfig}}
\vskip 10pt


Figure \ref{histogramfig} shows the distribution of galaxies
in bins of fractional difference between the infrared and optical
measurements.  The dispersions measured with the CO bandhead
are as much as 30\% and 40\% lower than their 
optical counterpart measurements, 
with the median difference being 11\% lower.
This is in the opposite sense to the predictions of \citet{bae02}.
Those authors's models show that dust decreases the
velocity dispersion in central regions
for radial and isotropic galaxies.  Only for their
tangential orbital structures does dust increase the central
dispersion, but early-type galaxies are not usually thought to
be dominated by tangential orbits.  Also, the magnitude of this
effect seems to be larger than predicted by \citet{bae02}.
They predict
an effect of a few percent for modest amounts of dust, and these
results show an effect of up to 30\% for many galaxies.
\citet{bae02} do find an effect similar to these results (a
signficant decrease in the velocity dispersion) due to dust
scattering, but only at several effective radii.

\subsection{Dust, Equivalent Widths, and Template Make-up}

In Figure \ref{dustfig}, we examine the relationship between the
fractional difference between the two measurements of the dispersion
and the dust mass calculated from IRAS flux densities. 
If it is dust which is causing the difference
between the infrared and optical dispersions, then surely
it is the relative amount of dust in a galaxy, not the absolute
amount, which is important.  To examine this, we plot the fractional
difference versus the ratio of IRAS dust mass to $B$-band total
luminosity, as quoted in RC3.  This gives a rough estimate of
the relative importance of dust for each galaxy.  Using a $K$-band
magnitude would be better here because dust attenuation affects the
$B$ band strongly, but some of the galaxies in our sample do not have
near-IR photometric measurements in the same filter system, or even
at all.


\psfig{file=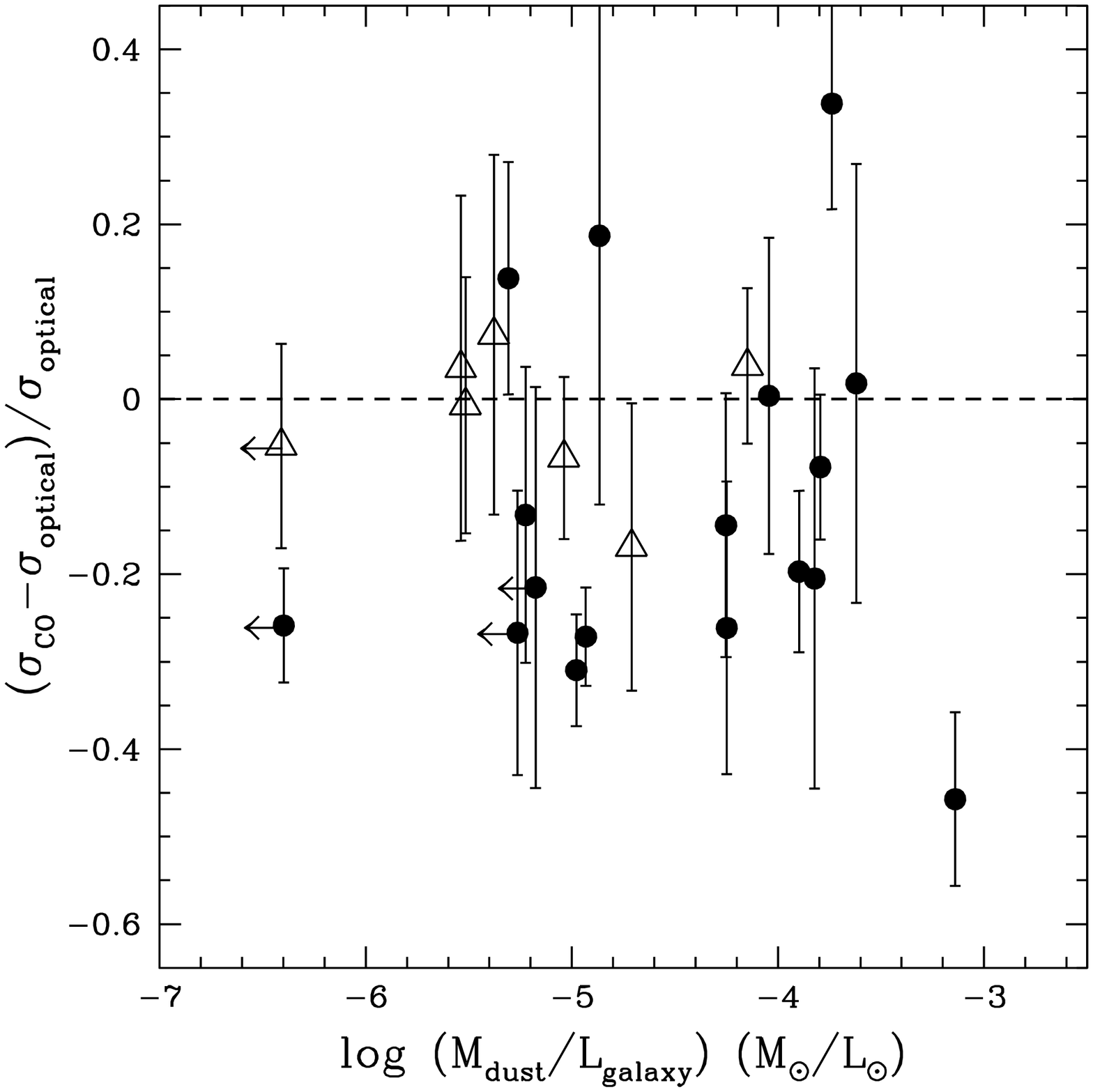,width=8.5cm,angle=0}
\figcaption[dust.eps]{Fractional difference between infrared
and optical dispersions as a function of the
ratio of dust mass to $B$-band luminosity.
The arrows indicate upper limits on the dust mass-to-light.
The dashed line represents equality between the
infrared and optical measurements.
\label{dustfig}}
\vskip 10pt



\begin{figure*}[b]
\centerline{\psfig{file=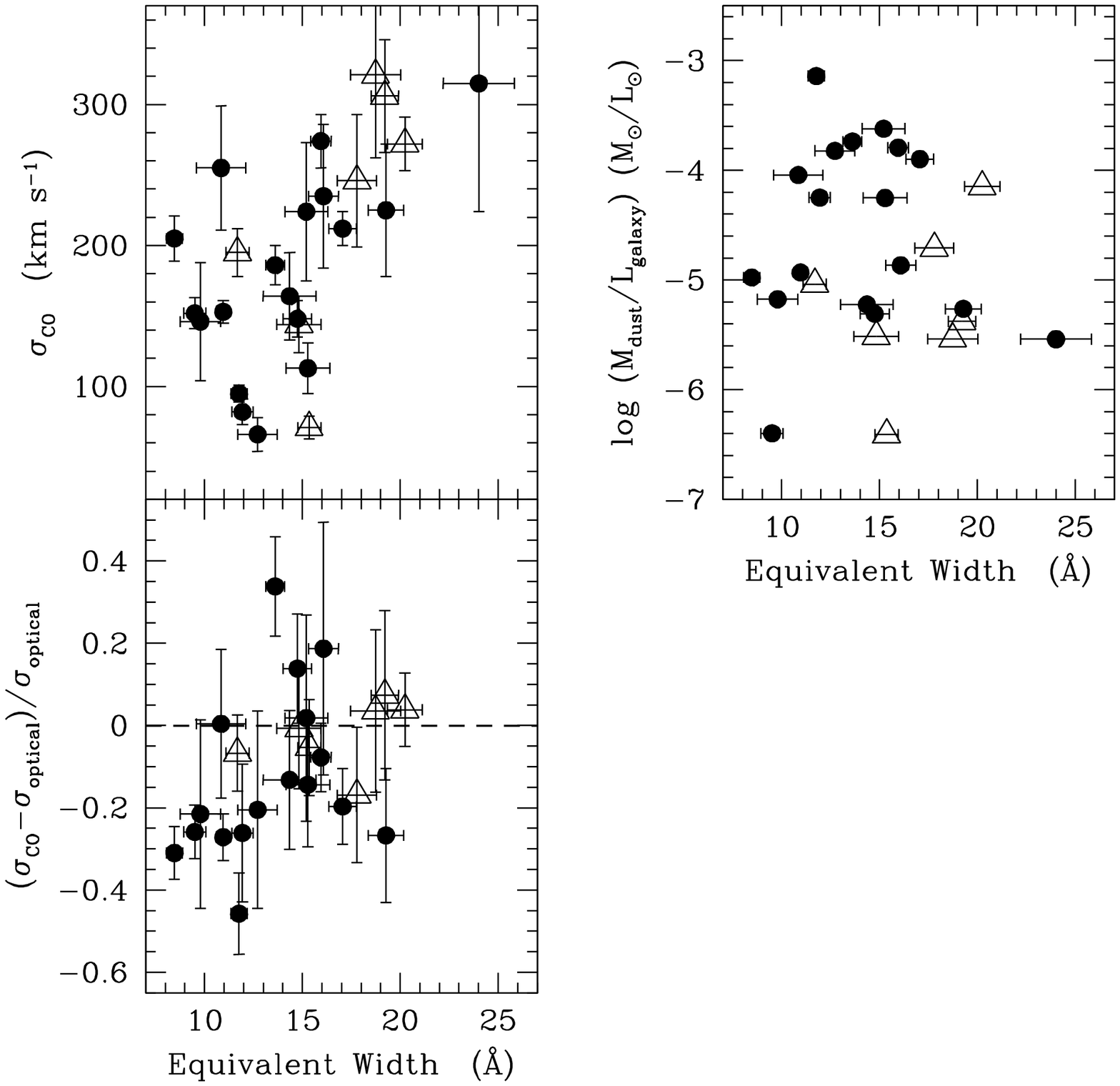,width=15cm,angle=0}}
\figcaption[equivwidth.eps]{Properties of the sample 
galaxies as they relate to the
measured CO bandhead equivalent width.  Shown are the relationships
between equivalent width and CO bandhead dispersion, IRAS dust mass,
and fractional difference between infrared and optical dispersions.
\label{EWfig}}
\end{figure*}



\begin{figure*}[b]
\centerline{\psfig{file=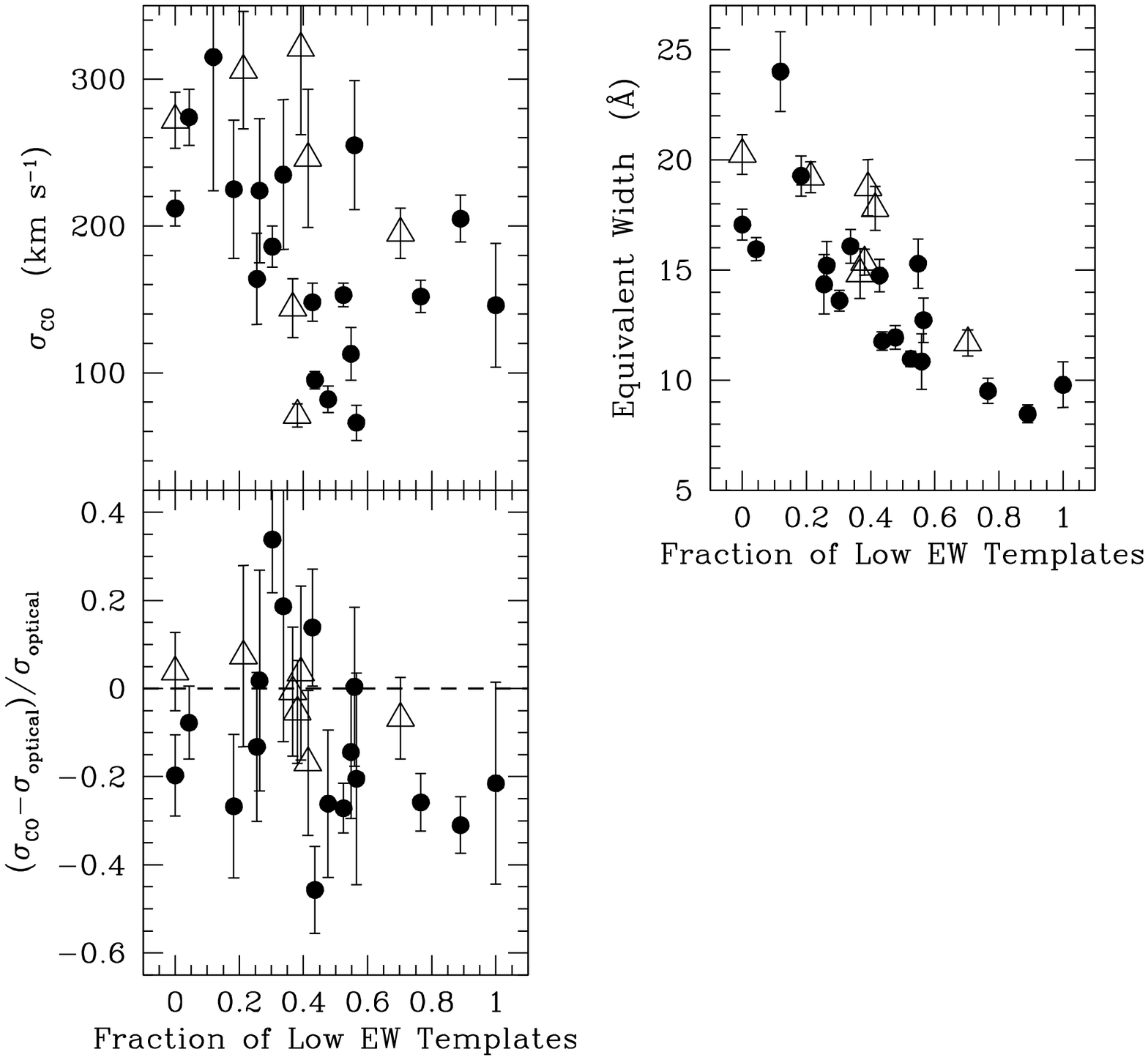,width=15cm,angle=0}}
\figcaption[fracpop.eps]{Same as Fig. \ref{EWfig}, but for the fractional
weight given to low equivalent width templates.
\label{fracfig}}
\end{figure*}


We calculate the Spearman rank-order correlation coefficient ($r_{s}$)
and the probability ($P_{r_{s}}$) to examine any statistical correlation here.
$P_{r_{s}}$ = 1 indicates that the data are completely uncorrelated,
while $P_{r_{s}}$ = 0 indicates complete correlation; the sign of 
$r_{s}$ indicates correlation or anticorrelation. 
For the whole sample, $P_{r_{s}}$ = 0.985 and $r_{s}$ = -0.0045;
for the sample minus the four galaxies for which the dust mass is
an upper limit, $P_{r_{s}}$ = 0.369 and $r_{s}$ = -0.212.
These results imply that the dust mass-to-light ratio and the fractional
difference are not likely to be correlated.  There is a slight hint of
an anticorrelation in the sample minus the excluded galaxies,
but the evidence is not strong.
The dust may be important in these differences, but there must
be other effects at work besides the relative amount of dust
in a galaxy.  According to the models
described earlier, the effect of dust depends on the orbital structure,
which could vary for the galaxies in this sample. The radial dust
distribution also will change these effects and may be important
in these fractional differences.

In addition, stellar population 
differences between the two wavelength regimes could cause such
an effect.  If the $V-K$ colors of the galaxies were bluer toward
their centers, the $K$-band light would sample less central, lower-velocity
regions of the galaxy and the $K$-band dispersion would be lower, as
found here. To test this, we are measuring $K$-band surface 
photometry of these galaxies using the recently available data
of the {\it Two Micron All Sky Survey} ({\it 2MASS}).
This surface photometry has already been obtained for one galaxy in our 
sample, which has good agreement between its infrared and 
optical dispersions: M32. \citet{pel93} reports surface
photometry for several optical and infrared colors and finds that
M32's profile is practically flat in all colors, including the
optical-infrared colors.  This is consistent with negligible 
dust and matching optical and infrared kinematics, but does not
help us understand the systematic differences between optical
and infrared dispersions.  Early-type galaxies are redder
in their centers in optical colors, which, if also true
in optical-infrared colors, would result in a change in 
dispersion opposite to what we find.

We can examine the relationship between equivalent width and
these quantities.  Table \ref{tbl-3} shows the measured equivalent
widths for the sample galaxies.  In stars, the CO bandhead
equivalent width increases from warmer, younger stars to cooler,
older stars, with a typical early K giant having an equivalent
width on the order of $\sim10$ \AA~\citep{kle86,sch00}.
Figure \ref{EWfig} shows the distribution of equivalent widths
for this sample and how the sample properties depend on this
quantity.  The measured CO bandhead dispersion seems to be
correlated with equivalent width, which is reasonable
considering other known metallicity-dispersion relations.
The probability $P_{r_{s}}$ = 0.036 indicates a high
probability of a correlation ($r_{s}$ = 0.438).
In addition, there appears to be
correlation between the measured equivalent width of the
galaxy and the fractional difference between the two
dispersion measurements.  For these quantities, $P_{r_{s}}$ = 0.054
and $r_{s}$ = 0.436.


\begin{figure*}[b]
\centerline{\psfig{file=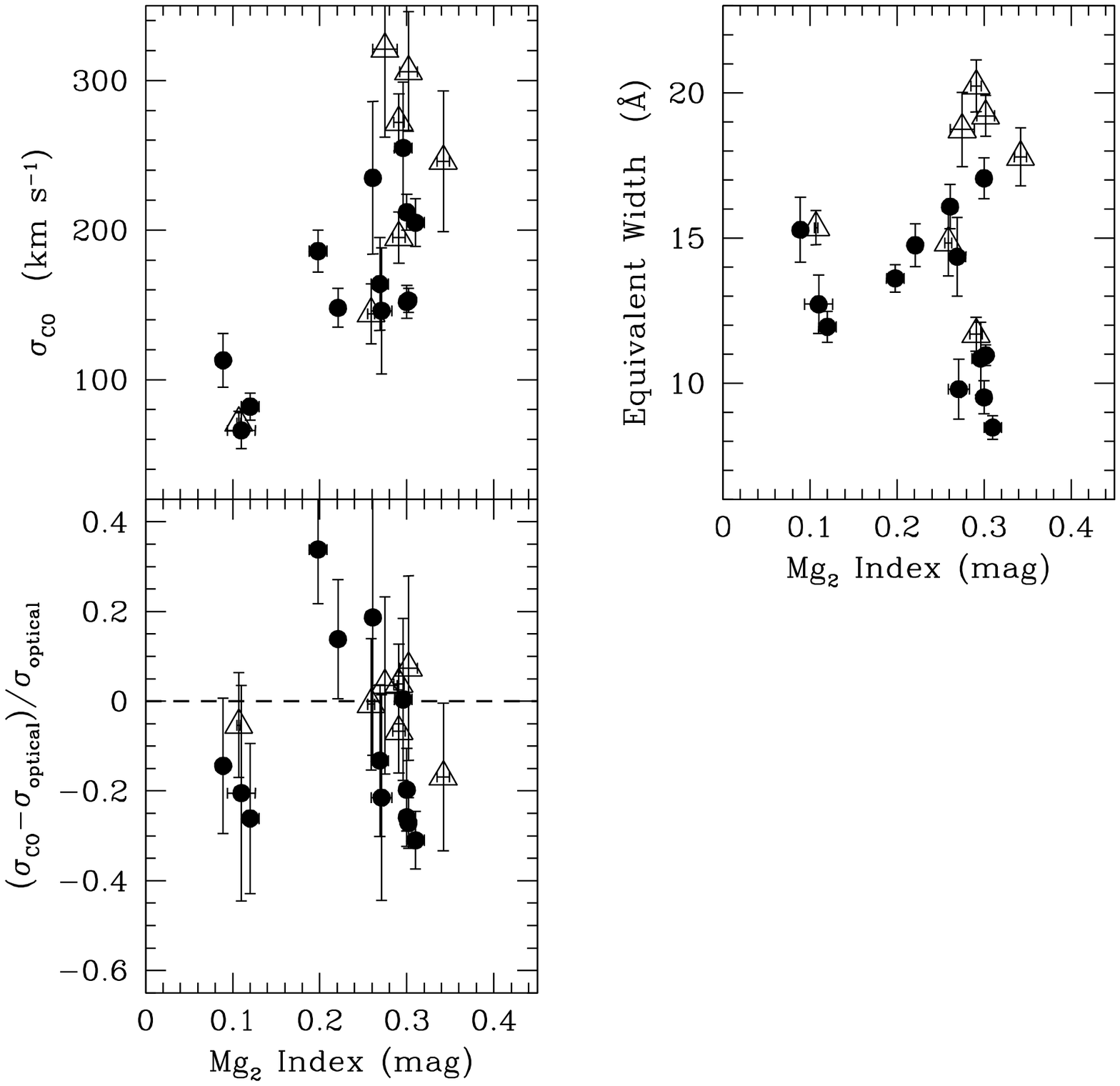,width=15cm,angle=0}}
\figcaption[Mgindex.eps]{Properties of the sample galaxies as they relate
to the Mg$_{2}$ index (from the literature).  Shown are the
Mg$_{2}$ index versus CO bandhead dispersion, CO bandhead
equivalent width, and fractional difference between infrared
and optical dispersions.
\label{Mgfig}}
\end{figure*}


Also interesting is the
distribution of equivalent widths of the galaxies.  Almost all of
the galaxies have equivalent widths over 10 \AA, to over 20 \AA.  
K giants have equivalent widths between 10 and 15 \AA, and the only stars
with equivalent widths higher than about 15 \AA~are M giants and
K and M supergiants.  The stellar population in the infrared must
have a large fraction of very cool stars to 
match the measured equivalent widths. According to these data, the light
at these wavelengths is largely dominated by M stars (probably giants).
This was suggested by
\citet{rie79} decades ago based on near-IR photometric measurements,
and provides important information about where the light is coming from
at these wavelengths. By number, the effect is not as extreme.  A typical
M star (dwarf, giant, or supergiant) is 6 or 7 times brighter than its
K star counterpart in the $K$ band, so while finding the light dominated
by M giants does imply that there is a significant population there,
there is still room for other stellar types.  This information must be taken
into account when measuring kinematics in the near-infrared.  Previous work
has often used K giants for stellar templates, which easily could be the
wrong choice and will effect the derived kinematics in a template-sensitive
feature like the CO bandhead.

We also have information about the relative weights
given to the different template stars by the fitting program.  The
LOSVD fitting is free to choose from eight different template stars
to achieve the best fit.  The different galaxies usually use only two
or three of these stars.  Figure \ref{fracfig} shows how these weights
are related to other galaxy characteristics.  We quantify the make-up of
the average stellar template chosen by the fitting program by the fraction
of the weight given to low equivalent width (low-EW) stars.  These are the
template stars with an equivalent width below 10 \AA; moderate changes in 
the choice of cut-off
do not affect the results.  The galaxies
in our sample vary from giving all the weight to low-EW
stars to using none of these stars.  The correlation between
the equivalent width of the galaxy and the weight given to templates of
a certain equivalent width (upper right-hand quadrant) 
is expected; galaxies with high equivalent widths
require templates of high equivalent width to achieve a good fit. The
apparent correlation between the fractional difference between optical
and IR dispersions and the fraction of low-EW templates used 
(upper left-hand quadrant) is
just that expected from the previously mentioned correlation
and the relationship between galaxy equivalent width and dispersion
differences.

Different spectral indices have been used to study the
chemical history of galaxies.  The Mg$_{2}$ index is perhaps
the most widely studied for this purpose; it is a measurement
of the flux deficit in the lines compared to the neighboring
continuum. Comparing this index with our infrared data may
give us useful information on these aspects of galaxies.
Twenty of the galaxies in our sample have Mg$_{2}$ indices in
the compilation of \citet{gol98} and the updates to that
compilation available on HYPERCAT 
(\url{http://www-obs.univ-lyon1.fr/hypercat/}).
This catalog of published absorption-line Mg$_{2}$ indices has
been zero-point corrected and transformed to a homogeneous
system. This facilitates comparison between different authors'
measurements.  We used measurements from \citet{tra98} when
available; table \ref{tbl-3} shows the Mg$_{2}$ indices used and
their sources in the literature.  The actual values are taken from
HYPERCAT where they have been transformed to a standard system.
Figure \ref{Mgfig} illustrates the relationships between the
Mg$_{2}$ index and the CO bandhead equivalent width, our velocity 
dispersion measurement, and the fractional difference between
the optical and infrared dispersions. The Mg$_{2}$ index appears to
be tightly correlated with velocity dispersion ($P_{r_{s}}$ = 0.105,
$r_{s}$ = 0.383); this is the familiar and well-studied Mg-$\sigma$
relation.  Further investigation into the scatter, slope, etc.~of this
relation using infrared velocity dispersion measurements (rather than
optical, as is currently done) may reveal important information about
this scaling relation. The CO bandhead equivalent
width is not very likely correlated with the Mg$_{2}$ index ($P_{r_{s}}$ = 
0.290, $r_{s}$ = -0.249).  This is perhaps surprising because 
both the CO equivalent width and the 
Mg$_{2}$ index appear to be significantly correlated with the 
velocity dispersion, which could cause them to be correlated with
each other. We do not see such a relationship, which could be due 
to our small sample size.

\subsection{Messier 32}

Messier 32 (NGC 221) is an important galaxy which is part of this sample.
It is the compact elliptical companion of the Andromeda Galaxy here in the
Local Group, and is extremely well-studied.  It was not detected
by IRAS and has an upper limit on the dust mass of less than 100 
M$_{\odot}$.  Thus, it should be an excellent candidate for calibrating
this technique and exploring the case of minimal dust.  It is important
to keep in mind, however, that in the infrared we are looking at
what may be a different stellar population with perhaps different
kinematic properties.  For example, if the infrared population is
more centrally concentrated relative to the whole potential than the
optical population, the infrared
dispersion would be higher than the optical dispersion because it samples
more central, higher-velocity stars.

Unfortunately,
there has been some observational difficulty with this galaxy.  It is
at nearly zero redshift ($cz = -145$ km s$^{-1}$) and thus the CO
bandhead falls exactly in a region of troublesome variability in
atmospheric absorption.  This feature was not well removed by the
flattening process described above.  The main reason is that NGC 221 is
far brighter than any other galaxy in this sample, and thus its total
integration time was less than 30 minutes.  Such a short integration
time made this galaxy vulnerable to problems with sky fluctuations.
The other galaxies in the sample required longer exposure times which
allowed these sky fluctuations to average out.  Also, this particular 
atmospheric feature is not a problem for galaxies with any considerable 
redshift, as the feature is moved away from this region.  The problem
area is visible in Figure \ref{gal1}; 
it is the ``bump'' between 2.296 and 2.298 $\mu$m.
This region was ignored during the LOSVD fitting (a happy benefit of
fitting the LOSVD directly in pixel space) and is located on the
red edge (rather than the sharp blue edge which dominates the fit)
but it does make it worrisome
to make strong statements based on the
results for this galaxy.  Despite these observational difficulties,
we measure a velocity disperion
of 71 $\pm$ 8 km s$^{-1}$, which is very close to the value of
75 km s$^{-1}$ from \citet{vdm94}. For the equivalent width
of this galaxy, we choose to use the equivalent width of the fit
made by the LOSVD extraction rather than of the observed spectrum,
as explained in a previous section.  This removes the rather large
effect ($\sim$20\%) of the sky feature from the equivalent width
measurement.

\vskip 30pt
\section{Conclusions}

In this project, we have observed 25 nearby early-type galaxies
and measured their stellar kinematics using the
$2.29 \mu$m (2-0) $^{12}$CO absorption bandhead.  We compare
the infrared velocity dispersions of these galaxies to optical
dispersions from the literature and find that the IR dispersions
are generally shifted to lower values relative to the optical dispersions,
between 5 and 30\% for most galaxies. However, this effect is mainly
driven by the S0 galaxies; pure ellipticals tend to have
nearly zero offset, on average. If dust in S0 galaxies exists
mainly in a disk which is cospatial with the stellar disk, optical
dispersions will be biased against measuring light from this cold
component and hence will come from the hot bulge component. This
may explain the effect we see, since we find lower CO dispersions
for lenticular galaxies, but it is currently unclear how dust
is distributed in galaxies.  More data is needed on the dust content
and distribution of galaxies to fully understand this.

We have calculated the dust masses implied by IRAS flux densities for
these galaxies and examined the relationship between the differences
in optical versus infrared dispersions and the amount of dust in
a galaxy. We do not find a strong relationship, but dust may still
be an important contributor to these differences.  Both the amount
and distribution of the dust can affect any differences in the mean
dispersion. We have also
calculated the equivalent width of this feature and compared it
to other galaxy properties.  The equivalent widths of the
galaxies are quite high, indicating that
the light is dominated by very cool (i.e. M) stars.

There are many extensions to be made from this first study.
Because of the technique used to deconvolve the galaxy spectra, we are
able to exploit the full LOSVDs, not just the dispersions.  There
is more kinematic information in the fits to these data than just
the width of the LOSVD.  Future work will be able to utilize
the higher moments of the LOSVD to examine more detailed kinematics
of sample galaxies.  These higher moments are helpful to understand
the orbits in galaxies.  Another important next step in 
this analysis will be to combine kinematic
information with infrared photometry.  {\it 2MASS}
has already observed most of the galaxies in this early-type
sample, as well as the late-type galaxies mentioned above, and
makes these images freely available.  We can use this photometry, in
conjunction with optical images, to constrain infrared vs. optical
stellar population differences and dust distributions.  Also, we can
construct an all-infrared Fundamental Plane, utilizing kinematic
and photometric $K$-band data, to explore regularity in galaxy
formation and evolution.

\acknowledgments{We would 
like to thank Dan Lester for his help and input as we have
begun using CoolSpec. We gratefully acknowledge McDonald Observatory.
This research project is supported by the Texas Advanced Research
Program and Grant No. 003658-0243-2001.}

\clearpage
\begin{deluxetable}{clrcrr}
\tablecaption{Basic properties of sample galaxies \label{tbl-1}}
\tablewidth{0pt}
\tablehead{
\colhead{Galaxy} & \colhead{Type} & \colhead{$v$} &
\colhead{$(m-M)$} & \colhead{$D$} & \colhead{$M_B$} \\
\colhead{} & & (km s$^{-1}$) & &(Mpc) & (mag)
}
\startdata

NGC 0221 &    cE2         & -145  &  24.55  &   0.813   &  -15.52  \\   
NGC 0315 &    E (LINER)   & 4942  &  --     &   70.600  &  -22.04  \\  
NGC 0821 &    E6          & 1735  &  31.91  &   24.099  &  -20.24  \\  
NGC 0984 &    S0          & 4352  &  --     &   62.171  &  -20.17  \\  
NGC 1023 &    S0          &  637  &  30.29  &   11.429  &  -19.94  \\ 
NGC 1161 &    S0          & 1954  &  --     &   27.914  &  -20.18  \\  
NGC 1400 &    S0          &  558  &  32.11  &   26.424  &  -20.19  \\  
NGC 1407 &    E0          & 1779  &  32.30  &   28.840  &  -21.59  \\  
NGC 2110 &    S0 (Sy2)    & 2335  &  --     &   33.357  &  -18.62  \\  
NGC 2293 &    S0 pec      & 2037  &  31.16  &   17.061  &  -18.88  \\ 
NGC 2380 &    S0          & 1782  &  32.05  &   25.704  &  -19.78  \\ 
NGC 2681 &    S0/a        &  692  &  31.18  &   17.219  &  -20.09  \\ 
NGC 2768 &    S0          & 1373  &  31.75  &   22.387  &  -20.91  \\  
NGC 2787 &    S0 (LINER)  &  696  &  29.37  &   7.482   &  -17.55  \\  
NGC 2974 &    E4          & 2072  &  31.66  &   21.478  &  -19.79  \\ 
NGC 3377 &    E5-6        &  665  &  30.25  &   11.220  &  -19.01  \\ 
NGC 3998 &    S0 (Sy1)    & 1040  &  30.75  &   14.125  &  -19.14  \\ 
NGC 4150 &    S0          &  226  &  30.69  &   13.740  &  -18.25  \\ 
NGC 5195 &    S0 pec (LINER)&  465  &   29.42  &   7.656    & -18.97   \\ 
NGC 5866 &    S0 (LINER)    &  672  &   30.93  &   15.346   & -20.19   \\ 
NGC 6548 &    S0            & 2174  &   31.81  &   23.014   & -19.08   \\ 
NGC 6703 &    S0            & 2461  &   32.13  &   26.669   &  -19.81  \\ 
NGC 7332 &    S0 pec        & 1172  &   31.81  &   23.014   &  -19.79  \\ 
NGC 7619 &    E             & 3762  &   33.62  &   52.966   &  -21.52  \\ 
NGC 7743 &    S0+ (Sy2)     & 1710  &   31.58  &   20.701   &  -19.20  \\

\enddata
\end{deluxetable}

\clearpage
\begin{deluxetable}{cccccc}
\tablecaption{IRAS fluxes and dust characteristics of sample galaxies 
\label{tbl-2}}
\tablewidth{0pt}
\tablehead{
\colhead{Galaxy} & 
\colhead{$S(60)$}& \colhead{$S(100)$} &
\colhead{$T_d$} & \colhead{log $M_d$}& \colhead{log ($M_d/L_{galaxy,B}$)} \\
\colhead{} & (mJy) & (mJy) & (K) & (M$_{\odot}$) & (M$_{\odot}$/L$_{\odot}$)
}
\startdata

NGC 0221 &    $<$85 &  $<$1412 &  -- & $<$1.99  & $<$-6.41   \\  
NGC 0315 &    310 &    400 &   42.3  &    5.47  &  -5.54   \\  
NGC 0821 &    $<$41 &    500 &   --  &    5.25  &  -5.04   \\  
NGC 0984 &    140 &    140 &   48.0  &    4.72  &  -5.54   \\  
NGC 1023 &    $<$32 &    $<$75 &   --& $<$3.77  & $<$-6.40  \\  
NGC 1161 &   1945 &  $<$9358 &   --  &    6.47  &  -3.79  \\  
NGC 1400 &    740 &   3280 &   27.2  &    6.37  &  -3.90  \\  
NGC 1407 &    140 &    480 &   29.2  &    5.45  &  -5.38   \\  
NGC 2110 &   4129 &   5676 &   41.1  &    6.02  &  -3.62   \\  
NGC 2293 &    380 &   2810 &   --    &    5.70  &  -4.04   \\  
NGC 2380 &     60 &   $<$388 &   --  &    4.88  &  -5.22  \\  
NGC 2681 &   6186 &  11770 &   35.9  &    5.98  &  -4.25  \\  
NGC 2768 &    390 &   1370 &   29.1  &    5.69  &  -4.87  \\  
NGC 2787 &    600 &   1180 &   35.5  &    4.28  &  -4.93  \\  
NGC 2974 &    420 &   1900 &   27.0  &    5.96  &  -4.15  \\  
NGC 3377 &    140 &    350 &   32.4  &    4.28  &  -5.52  \\  
NGC 3998 &    550 &   1150 &   34.6  &    4.87  &  -4.98  \\  
NGC 4150 &   1220 &   2670 &   34.1  &    5.24  &  -4.25  \\  
NGC 5195 &  38010 &   $<$511 &   --  &    6.64  &  -3.14  \\    
NGC 5866 &   5070 &  18680 &   28.7  &    6.53  &  -3.74  \\   
NGC 6548 &    $<$35 &   $<$119 &  -- & $<$4.56  & $<$-5.26  \\  
NGC 6703 &    $<$66 &   $<$200 &  -- & $<$4.94  & $<$-5.18   \\    
NGC 7332 &    210 &    410 &   35.5  &    4.80  &  -5.31   \\   
NGC 7619 &    $<$37 &    710 &   --  &    6.09  &  -4.71   \\   
NGC 7743 &    920 &   3400 &   28.7  &    6.05  &  -3.82   \\    

\enddata
\end{deluxetable}

\clearpage
\begin{deluxetable}{cccccccc}
\tablecaption{Derived quantities from observed spectra and
properties gathered from the literature\label{tbl-3}}
\tablewidth{0pt}
\tablehead{
\colhead{Galaxy} & 
\colhead{Total Exposure} &
\colhead{S/N} &
\colhead{Extraction} &
\colhead{Equivalent}& \colhead{$\sigma_{co}$} &
\colhead{$\sigma_{optical}$}  &
\colhead{Mg$_{2}$ Index} \\
\colhead{} & \colhead{Time (minutes)}& (pixel$^{-1}$)& Window & 
Width (\AA) & (km s$^{-1}$) & (km s$^{-1}$) & (mag)
}
\startdata

NGC 0221 &  24 & 27 &1.8$^{\prime\prime}$ $\times$ 14.0$^{\prime\prime}$& 15.4 $\pm$ 0.59  &  71 $\pm$  8 &  75 $\pm$  4\tablenotemark{a} &  0.107 $\pm$ 0.002\tablenotemark{b} \\
NGC 0315 & 160 & 13 &1.8$^{\prime\prime}$ $\times$ 15.4$^{\prime\prime}$& 18.7 $\pm$ 1.3   & 321 $\pm$ 59 & 310 $\pm$ 16\tablenotemark{b} &  0.275 $\pm$ 0.014\tablenotemark{b} \\
NGC 0821 &  96 & 30 &1.8$^{\prime\prime}$ $\times$ 11.9$^{\prime\prime}$& 11.7 $\pm$ 0.58  & 195 $\pm$ 17 & 209 $\pm$ 10\tablenotemark{c} &  0.291 $\pm$ 0.007\tablenotemark{b} \\
NGC 0984 & 168 &  8 &1.8$^{\prime\prime}$ $\times$ 14.7$^{\prime\prime}$& 24.0 $\pm$ 1.8   & 315 $\pm$ 91 &  --      &   --\\

NGC 1023 &  48 & 28 &1.8$^{\prime\prime}$ $\times$ 17.5$^{\prime\prime}$&  9.5 $\pm$ 0.57  & 152 $\pm$ 11 & 205 $\pm$ 10\tablenotemark{d} &  0.340 $\pm$ 0.005\tablenotemark{b} \\
NGC 1161 &  72 & 32 &1.8$^{\prime\prime}$ $\times$ 14.0$^{\prime\prime}$& 15.9 $\pm$ 0.52  & 274 $\pm$ 19 & 297 $\pm$ 17\tablenotemark{e} &  -- \\  
NGC 1400 &  40 & 26 &1.8$^{\prime\prime}$ $\times$ 10.5$^{\prime\prime}$& 17.1 $\pm$ 0.70  & 212 $\pm$ 12 & 264 $\pm$ 26\tablenotemark{b} &  0.300 $\pm$ 0.007\tablenotemark{b} \\ 
NGC 1407 & 240 & 29 &1.8$^{\prime\prime}$ $\times$ 17.5$^{\prime\prime}$& 19.2 $\pm$ 0.70  & 306 $\pm$ 40 & 285 $\pm$ 40\tablenotemark{b} &  0.302 $\pm$ 0.010\tablenotemark{b} \\
NGC 2110 & 120 & 12 &1.8$^{\prime\prime}$ $\times$ 12.6$^{\prime\prime}$& 15.2 $\pm$ 1.1  &  224 $\pm$ 49 & 220 $\pm$ 25\tablenotemark{f} &  -- \\ 
NGC 2293 & 120 & 14 &1.8$^{\prime\prime}$ $\times$ 14.0$^{\prime\prime}$& 10.8 $\pm$ 1.3  &  255 $\pm$ 44 & 254 $\pm$ 13\tablenotemark{g} &  0.296 $\pm$ 0.010\tablenotemark{m} \\
NGC 2380 &  96 & 18 &1.8$^{\prime\prime}$ $\times$ 10.5$^{\prime\prime}$& 14.4 $\pm$ 1.4  &  164 $\pm$ 31 & 189 $\pm$  9\tablenotemark{h} &  0.269 $\pm$ 0.010\tablenotemark{m} \\  
NGC 2681 & 120 & 31 &1.8$^{\prime\prime}$ $\times$ 17.5$^{\prime\prime}$& 11.9 $\pm$ 0.54 &   82 $\pm$  9 & 111 $\pm$ 22\tablenotemark{b} &  0.120 $\pm$ 0.010\tablenotemark{n} \\
NGC 2768 & 120 &  7 &1.8$^{\prime\prime}$ $\times$ 11.2$^{\prime\prime}$& 16.0 $\pm$ 0.77 &  235 $\pm$ 51 & 198 $\pm$ 28\tablenotemark{b} &  0.261 $\pm$ 0.006\tablenotemark{b} \\
NGC 2787 &  72 & 48 &1.8$^{\prime\prime}$ $\times$ 17.5$^{\prime\prime}$& 11.0 $\pm$ 0.35 &  153 $\pm$  8 & 210 $\pm$ 12\tablenotemark{b} &  0.302 $\pm$ 0.007\tablenotemark{b} \\
NGC 2974 & 144 & 21 &1.8$^{\prime\prime}$ $\times$ 14.0$^{\prime\prime}$& 20.2 $\pm$ 0.89 &  272 $\pm$ 19 & 262 $\pm$ 13\tablenotemark{g} &  0.291 $\pm$ 0.006\tablenotemark{b} \\
NGC 3377 & 166 & 10 &1.8$^{\prime\prime}$ $\times$ 10.5$^{\prime\prime}$& 14.8 $\pm$ 1.1  &  144 $\pm$ 20 & 145 $\pm$  7\tablenotemark{i} &  0.259 $\pm$ 0.004\tablenotemark{b} \\  
NGC 3998 &  72 & 39 &1.8$^{\prime\prime}$ $\times$ 10.5$^{\prime\prime}$&  8.5 $\pm$ 0.41 &  205 $\pm$ 16 & 297 $\pm$ 15\tablenotemark{j} &  0.310 $\pm$ 0.010\tablenotemark{o} \\
NGC 4150 &  72 &  8 &1.8$^{\prime\prime}$ $\times$ 7.0$^{\prime\prime}$& 15.3 $\pm$ 1.1  &  113 $\pm$ 18 & 132 $\pm$ 10\tablenotemark{k} &  0.089 $\pm$ 0.005\tablenotemark{b} \\
NGC 5195 &  72 & 35 &1.8$^{\prime\prime}$ $\times$ 7.0$^{\prime\prime}$& 11.8 $\pm$ 0.42 &   95 $\pm$ 6  & 175 $\pm$ 30\tablenotemark{l} &   --  \\ 
NGC 5866 & 128 & 40 &1.8$^{\prime\prime}$ $\times$ 14.0$^{\prime\prime}$& 13.6 $\pm$ 0.47 &  186 $\pm$ 14 & 139 $\pm$ 7\tablenotemark{j} &  0.198 $\pm$ 0.010\tablenotemark{o} \\
NGC 6548 & 296 & 13 &1.8$^{\prime\prime}$ $\times$ 6.3$^{\prime\prime}$& 19.3 $\pm$ 0.91 &  225 $\pm$ 47 & 307 $\pm$ 23\tablenotemark{e} &  --  \\    
NGC 6703 & 208 &  7 &1.8$^{\prime\prime}$ $\times$ 9.1$^{\prime\prime}$&  9.8 $\pm$ 1.0  &  146 $\pm$ 42 & 186 $\pm$  9\tablenotemark{j} &  0.271 $\pm$ 0.012\tablenotemark{b} \\
NGC 7332 &  96 & 18 &1.8$^{\prime\prime}$ $\times$ 6.3$^{\prime\prime}$& 17.8 $\pm$ 0.73 &  148 $\pm$ 13 & 130 $\pm$ 10\tablenotemark{b} &  0.221 $\pm$ 0.007\tablenotemark{b} \\
NGC 7619 &  96 & 12 &1.8$^{\prime\prime}$ $\times$ 14.0$^{\prime\prime}$& 17.8 $\pm$ 1.0  &  246 $\pm$ 47 & 296 $\pm$ 15\tablenotemark{g} &  0.342 $\pm$ 0.007\tablenotemark{b} \\
NGC 7743 & 120 & 18 &1.8$^{\prime\prime}$ $\times$ 4.2$^{\prime\prime}$& 12.7 $\pm$ 1.0  &   66 $\pm$ 12 &  83 $\pm$ 20\tablenotemark{f} &  0.110 $\pm$ 0.016\tablenotemark{p} \\

\enddata

\footnotesize{\tablerefs{
(a) \citet{vdm94};
(b) \citet{tra98};
(c) \citet{geb03};
(d) \citet{bow01};
(e) \citet{ton81};
(f) \citet{nel95};
(g) \citet{jor95};
(h) \citet{dav87};
(i) \citet{kor98};
(j) \citet{fis97};
(k) \citet{din95};
(l) \citet{whi83};
(m) \citet{fab89};
(n) \citet{bur88};
(o) \citet{fis96};
(p) \citet{huc96};
}}

\end{deluxetable}

\end{document}